\pdfoutput=1
\documentclass{article}

\usepackage{PRIMEarxiv}

\usepackage[utf8]{inputenc}
\usepackage[T1]{fontenc}
\usepackage{amssymb}
\usepackage{amsmath}
\usepackage{amsfonts}
\usepackage{booktabs}
\usepackage{multirow}
\usepackage{array}
\usepackage{tabularx}
\usepackage{float}
\usepackage{placeins}
\usepackage{nicefrac}
\usepackage{microtype}
\usepackage{graphicx}
\usepackage{xcolor}
\usepackage{fancyhdr}
\usepackage{xspace}
\xspaceaddexceptions{\%}
\usepackage[round]{natbib}
\usepackage{url}
\usepackage{xurl}
\usepackage{hyperref}
\hypersetup{colorlinks=true, linkcolor=blue, citecolor=blue, urlcolor=blue}
\hypersetup{pdftitle={You Know What You Will Like: People Predict Their Own Ratings Better Than Recommender Systems},
  pdfauthor={Zihan Chen, Lei Nico Zheng, Di Zhu},
  pdfkeywords={recommender systems, preference elicitation, collaborative filtering, human versus algorithmic prediction, affective forecasting, human-AI complementarity, dispersed knowledge, MovieLens}}
\graphicspath{{figures/}}

\newcommand{\AblGbmAllR}{0.70\xspace}
\newcommand{\AblGbmMachR}{0.56\xspace}
\newcommand{\AblGuessConc}{71\xspace}
\newcommand{\AblGuessR}{0.72\xspace}
\newcommand{\AblGuessRsq}{0.51\xspace}
\newcommand{\AblMachConc}{66\xspace}
\newcommand{\AblMachR}{0.59\xspace}
\newcommand{\AblMachRsq}{0.34\xspace}
\newcommand{\AblPersRsq}{0.49\xspace}
\newcommand{\AgeForgot}{23\xspace}
\newcommand{\AgeLater}{17\xspace}
\newcommand{\AskGainCorr}{0.08\xspace}
\newcommand{\AskHalf}{63\xspace}
\newcommand{\AskHalfCI}{41 to 84\xspace}
\newcommand{\AskQuarter}{36\xspace}
\newcommand{\AskQuarterCI}{18 to 57\xspace}
\newcommand{\BalCertRaw}{0.43\xspace}
\newcommand{\BalCertW}{0.19\xspace}
\newcommand{\BalFollowRaw}{0.33\xspace}
\newcommand{\BalFollowW}{$-$0.08\xspace}
\newcommand{\BalGuessRaw}{0.79\xspace}
\newcommand{\BalGuessW}{0.08\xspace}
\newcommand{\BalMovieRaw}{0.54\xspace}
\newcommand{\BalMovieW}{0.46\xspace}
\newcommand{\BiasAbsKN}{0.36\xspace}
\newcommand{\BiasAbsMF}{0.27\xspace}
\newcommand{\BiasAbsML}{0.51\xspace}
\newcommand{\BiasG}{$-$0.02\xspace}
\newcommand{\BiasValRMSE}{0.841\xspace}
\newcommand{\Boot}{2,000\xspace}
\newcommand{\BroadForgot}{86\xspace}
\newcommand{\BroadLater}{11\xspace}
\newcommand{\CalGFive}{4.5\xspace}
\newcommand{\CalGHalf}{0.7\xspace}
\newcommand{\CalGThree}{3.1\xspace}
\newcommand{\CalMLFive}{3.8\xspace}
\newcommand{\CalMLFourHalf}{4.0\xspace}
\newcommand{\CeilConc}{86\xspace}
\newcommand{\CeilExact}{49\xspace}
\newcommand{\CeilGapYears}{3.0\xspace}
\newcommand{\CeilR}{0.86\xspace}
\newcommand{\CertBiasFiveAbs}{0.40\xspace}
\newcommand{\CertBiasFiveSE}{0.20\xspace}
\newcommand{\CertExtreme}{0.40\xspace}
\newcommand{\ConcCIG}{65 to 78\xspace}
\newcommand{\ConcG}{72\xspace}
\newcommand{\ConcGMonth}{67\xspace}
\newcommand{\ConcGNear}{73\xspace}
\newcommand{\ConcGQuarter}{56\xspace}
\newcommand{\ConcGRec}{70\xspace}
\newcommand{\ConcGWeek}{70\xspace}
\newcommand{\ConcIM}{62\xspace}
\newcommand{\ConcKN}{67\xspace}
\newcommand{\ConcKNNear}{67\xspace}
\newcommand{\ConcKNQuarter}{73\xspace}
\newcommand{\ConcKNWeek}{65\xspace}
\newcommand{\ConcMF}{64\xspace}
\newcommand{\ConcMFMonth}{69\xspace}
\newcommand{\ConcMFNear}{61\xspace}
\newcommand{\ConcMFQuarter}{66\xspace}
\newcommand{\ConcMFSnap}{60\xspace}
\newcommand{\ConcMFWeek}{64\xspace}
\newcommand{\ConcML}{56\xspace}
\newcommand{\ConcMLNear}{56\xspace}
\newcommand{\ConcMLRec}{56\xspace}
\newcommand{\ConcMLWeek}{54\xspace}
\newcommand{\ConcMachMax}{67\xspace}
\newcommand{\ConcMembers}{77\xspace}
\newcommand{\ConcST}{66\xspace}
\newcommand{\ConcSTRec}{66\xspace}
\newcommand{\CopyEqual}{14.9\xspace}
\newcommand{\CopyShuffled}{13.1\xspace}
\newcommand{\DiffConcCIML}{5 to 27\xspace}
\newcommand{\DiffConcKN}{5\xspace}
\newcommand{\DiffConcMF}{8\xspace}
\newcommand{\DiffConcML}{16\xspace}
\newcommand{\DiffRBaseCIST}{0.09 to 0.26\xspace}
\newcommand{\DiffRBaseST}{0.17\xspace}
\newcommand{\DiffRCIMF}{0.05 to 0.26\xspace}
\newcommand{\DiffRCIML}{0.20 to 0.51\xspace}
\newcommand{\DiffRCIST}{0.03 to 0.18\xspace}
\newcommand{\DiffRMF}{0.15\xspace}
\newcommand{\DiffRML}{0.35\xspace}
\newcommand{\DiffRMovieCIMF}{0.07 to 0.22\xspace}
\newcommand{\DiffRMovieCIML}{0.24 to 0.45\xspace}
\newcommand{\DiffRST}{0.10\xspace}
\newcommand{\DiffSureCI}{0.07 to 0.32\xspace}
\newcommand{\DiffUnsureCI}{$-$0.13 to 0.54\xspace}
\newcommand{\FollowMax}{419\xspace}
\newcommand{\FollowMin}{5\xspace}
\newcommand{\ForgotYears}{4.2\xspace}
\newcommand{\FwdGain}{0.323\xspace}
\newcommand{\FwdGainCI}{0.124 to 0.531\xspace}
\newcommand{\FwdMSEGuess}{0.677\xspace}
\newcommand{\FwdMSEMach}{1.000\xspace}
\newcommand{\FwdSplit}{17 August 2023\xspace}
\newcommand{\FwdTest}{221\xspace}
\newcommand{\FwdTrain}{330\xspace}
\newcommand{\GapML}{1.17\xspace}
\newcommand{\GapOneStar}{48\xspace}
\newcommand{\GuessMachRsqPct}{31\xspace}
\newcommand{\HetCells}{12\xspace}
\newcommand{\HetGMax}{0.77\xspace}
\newcommand{\HetGMin}{0.58\xspace}
\newcommand{\HetLead}{11\xspace}
\newcommand{\HetMMax}{0.65\xspace}
\newcommand{\HetMMin}{0.41\xspace}
\newcommand{\HetSig}{7\xspace}
\newcommand{\HistCut}{9\xspace}
\newcommand{\HistFewGuess}{0.246\xspace}
\newcommand{\HistFewGuessCI}{0.135 to 0.373\xspace}
\newcommand{\HistFewTransfer}{0.068\xspace}
\newcommand{\HistFewTransferCI}{$-$0.006 to 0.148\xspace}
\newcommand{\HistMoreGuess}{0.116\xspace}
\newcommand{\HistMoreGuessCI}{$-$0.008 to 0.305\xspace}
\newcommand{\HistMoreTransfer}{0.002\xspace}
\newcommand{\HistMoreTransferCI}{$-$0.072 to 0.075\xspace}
\newcommand{\IPWDiffCIKN}{0.10 to 0.40\xspace}
\newcommand{\IPWG}{0.74\xspace}
\newcommand{\IPWKN}{0.50\xspace}
\newcommand{\IPWMF}{0.50\xspace}
\newcommand{\IPWML}{0.37\xspace}
\newcommand{\IPWTrimG}{0.74\xspace}
\newcommand{\IPWTrimMF}{0.51\xspace}
\newcommand{\IPWTrimML}{0.38\xspace}
\newcommand{\IPWess}{164\xspace}
\newcommand{\IPWwMax}{2.9\xspace}
\newcommand{\KnnBaseRMSE}{0.864\xspace}
\newcommand{\KnnK}{20\xspace}
\newcommand{\KnnShrink}{10\xspace}
\newcommand{\KnnValN}{5,000\xspace}
\newcommand{\KnnValRMSE}{0.779\xspace}
\newcommand{\KnownAdvMF}{0.12\xspace}
\newcommand{\LabelShare}{2.1\xspace}
\newcommand{\LagMedian}{58\xspace}
\newcommand{\LeadExtreme}{0.63\xspace}
\newcommand{\LeadExtremeSE}{0.32\xspace}
\newcommand{\LeadFive}{$-$0.62\xspace}
\newcommand{\LeadFiveSE}{0.81\xspace}
\newcommand{\LooMaxMF}{9\xspace}
\newcommand{\LooMaxML}{17\xspace}
\newcommand{\LooMinMF}{7\xspace}
\newcommand{\LooMinML}{14\xspace}
\newcommand{\MAEG}{0.61\xspace}
\newcommand{\MAEGFive}{0.48\xspace}
\newcommand{\MAEMFive}{0.99\xspace}
\newcommand{\MAEMachMin}{0.72\xspace}
\newcommand{\MAEcalG}{0.64\xspace}
\newcommand{\MAEcalMachMin}{0.72\xspace}
\newcommand{\MFFoldC}{1\xspace}
\newcommand{\MFSeedConcRange}{59 to 64\xspace}
\newcommand{\MFSeedRange}{0.53 to 0.54\xspace}
\newcommand{\MFValRMSE}{0.759\xspace}
\newcommand{\MFValRMSECut}{0.911\xspace}
\newcommand{\MFk}{64\xspace}
\newcommand{\NAfterTune}{437\xspace}
\newcommand{\NDupAccounts}{9,626\xspace}
\newcommand{\NDupRatings}{4.4~million\xspace}
\newcommand{\NExclPriorRating}{12\xspace}
\newcommand{\NExclPriorView}{7\xspace}
\newcommand{\NFive}{95\xspace}
\newcommand{\NForgot}{238\xspace}
\newcommand{\NForgotUsers}{108\xspace}
\newcommand{\NGuessMovies}{7,521\xspace}
\newcommand{\NGuessRetest}{540\xspace}
\newcommand{\NGuessUsers}{3,200\xspace}
\newcommand{\NGuesses}{26,020\xspace}
\newcommand{\NHHalf}{344\xspace}
\newcommand{\NHNinety}{297\xspace}
\newcommand{\NIPWMembers}{3,163\xspace}
\newcommand{\NIPWPop}{25,438\xspace}
\newcommand{\NLater}{551\xspace}
\newcommand{\NLaterMovies}{389\xspace}
\newcommand{\NLaterUsers}{277\xspace}
\newcommand{\NMonth}{355\xspace}
\newcommand{\NMonthMembers}{50\xspace}
\newcommand{\NMovieForgot}{18,886\xspace}
\newcommand{\NMovieLater}{9,455\xspace}
\newcommand{\NNewMem}{172\xspace}
\newcommand{\NNoExp}{409\xspace}
\newcommand{\NNotSeen}{28,462\xspace}
\newcommand{\NPromptUsers}{22,382\xspace}
\newcommand{\NPrompts}{3.0~million\xspace}
\newcommand{\NQuarterMembers}{27\xspace}
\newcommand{\NQuick}{284\xspace}
\newcommand{\NRec}{445\xspace}
\newcommand{\NReguess}{98\xspace}
\newcommand{\NRepeated}{1,462\xspace}
\newcommand{\NReprompt}{302\xspace}
\newcommand{\NRetest}{1,444\xspace}
\newcommand{\NRetestUsers}{615\xspace}
\newcommand{\NSure}{290\xspace}
\newcommand{\NTrainMax}{32.8~million\xspace}
\newcommand{\NTrainMin}{31.6~million\xspace}
\newcommand{\NTrainPool}{33.0~million\xspace}
\newcommand{\NUnsure}{87\xspace}
\newcommand{\NUserForgot}{1,941\xspace}
\newcommand{\NUserLater}{471\xspace}
\newcommand{\NearMembers}{52\xspace}
\newcommand{\NearPairs}{621\xspace}
\newcommand{\NewAdvMF}{0.57\xspace}
\newcommand{\NewDiffCIKN}{0.00 to 1.68\xspace}
\newcommand{\NewDiffCIMF}{$-$0.16 to 1.45\xspace}
\newcommand{\NewMemHistMedian}{101\xspace}
\newcommand{\NoAnswerShare}{98.8\xspace}
\newcommand{\OthMedian}{11\xspace}
\newcommand{\OthShare}{90\xspace}
\newcommand{\OvBalMax}{0.12\xspace}
\newcommand{\OvDiffCIKN}{0.06 to 0.36\xspace}
\newcommand{\OvDiffCIMF}{0.07 to 0.35\xspace}
\newcommand{\OvDiffCIML}{0.19 to 0.47\xspace}
\newcommand{\OvEss}{170\xspace}
\newcommand{\OvG}{0.75\xspace}
\newcommand{\OvKN}{0.54\xspace}
\newcommand{\OvMF}{0.53\xspace}
\newcommand{\OvML}{0.44\xspace}
\newcommand{\OvMembers}{2,812\xspace}
\newcommand{\OvMin}{195\xspace}
\newcommand{\OvMovies}{3,637\xspace}
\newcommand{\OvN}{16,490\xspace}
\newcommand{\OvRated}{496\xspace}
\newcommand{\PairCertMSE}{0.022\xspace}
\newcommand{\PairCertMSECI}{$-$0.001 to 0.046\xspace}
\newcommand{\PairGuessConcCI}{$-$3 to 13\xspace}
\newcommand{\PairGuessMSE}{0.236\xspace}
\newcommand{\PairGuessMSECI}{0.122 to 0.363\xspace}
\newcommand{\PairGuessMSEShare}{26\xspace}
\newcommand{\PairGuessTransferMSE}{0.183\xspace}
\newcommand{\PairGuessTransferMSECI}{0.085 to 0.295\xspace}
\newcommand{\PairIntMSE}{$-$0.008\xspace}
\newcommand{\PairIntMSECI}{$-$0.024 to 0.007\xspace}
\newcommand{\PairOthMSE}{0.014\xspace}
\newcommand{\PairOthMSECI}{$-$0.027 to 0.057\xspace}
\newcommand{\PairPersMSE}{0.049\xspace}
\newcommand{\PairPersMSECI}{$-$0.018 to 0.121\xspace}
\newcommand{\PairTransferMSE}{0.036\xspace}
\newcommand{\PairTransferMSECI}{$-$0.019 to 0.092\xspace}
\newcommand{\PromptEnd}{1 May 2024\xspace}
\newcommand{\PromptStart}{13 March 2023\xspace}
\newcommand{\QuickHalfHour}{202\xspace}
\newcommand{\RBaseG}{0.63\xspace}
\newcommand{\RBaseMachMax}{0.47\xspace}
\newcommand{\RG}{0.69\xspace}
\newcommand{\RGAfterTune}{0.68\xspace}
\newcommand{\RGBroad}{0.84\xspace}
\newcommand{\RGEarly}{0.69\xspace}
\newcommand{\RGEstab}{0.64\xspace}
\newcommand{\RGFam}{0.60\xspace}
\newcommand{\RGFewest}{0.65\xspace}
\newcommand{\RGForgot}{0.67\xspace}
\newcommand{\RGForgotYear}{0.79\xspace}
\newcommand{\RGHHalf}{0.66\xspace}
\newcommand{\RGHNinety}{0.71\xspace}
\newcommand{\RGHalf}{0.59\xspace}
\newcommand{\RGLate}{0.69\xspace}
\newcommand{\RGMonth}{0.69\xspace}
\newcommand{\RGNew}{0.63\xspace}
\newcommand{\RGNewMem}{0.76\xspace}
\newcommand{\RGNoExp}{0.71\xspace}
\newcommand{\RGNoReguess}{0.70\xspace}
\newcommand{\RGNoReprompt}{0.75\xspace}
\newcommand{\RGQuarter}{0.59\xspace}
\newcommand{\RGRec}{0.64\xspace}
\newcommand{\RGSure}{0.77\xspace}
\newcommand{\RGUnfam}{0.74\xspace}
\newcommand{\RGUnsure}{0.52\xspace}
\newcommand{\RGuessHat}{0.57\xspace}
\newcommand{\RGuessRetest}{0.80\xspace}
\newcommand{\RKN}{0.56\xspace}
\newcommand{\RKNEstab}{0.60\xspace}
\newcommand{\RKNForgot}{0.73\xspace}
\newcommand{\RKNHHalf}{0.58\xspace}
\newcommand{\RKNHNinety}{0.59\xspace}
\newcommand{\RKNNewMem}{0.52\xspace}
\newcommand{\RKNSnap}{0.47\xspace}
\newcommand{\RMF}{0.54\xspace}
\newcommand{\RMFAfterTune}{0.54\xspace}
\newcommand{\RMFEarly}{0.61\xspace}
\newcommand{\RMFEstab}{0.54\xspace}
\newcommand{\RMFForgot}{0.70\xspace}
\newcommand{\RMFHHalf}{0.54\xspace}
\newcommand{\RMFHNinety}{0.54\xspace}
\newcommand{\RMFHalf}{0.37\xspace}
\newcommand{\RMFLate}{0.47\xspace}
\newcommand{\RMFMonth}{0.55\xspace}
\newcommand{\RMFNewMem}{0.57\xspace}
\newcommand{\RMFNoExp}{0.53\xspace}
\newcommand{\RMFNoReguess}{0.59\xspace}
\newcommand{\RMFNoReprompt}{0.63\xspace}
\newcommand{\RMFQuarter}{0.50\xspace}
\newcommand{\RMFRec}{0.52\xspace}
\newcommand{\RMFSnap}{0.42\xspace}
\newcommand{\RMFam}{0.51\xspace}
\newcommand{\RMFewest}{0.41\xspace}
\newcommand{\RML}{0.34\xspace}
\newcommand{\RMLAfterTune}{0.37\xspace}
\newcommand{\RMLBroad}{0.54\xspace}
\newcommand{\RMLEstab}{0.31\xspace}
\newcommand{\RMLFewest}{0.13\xspace}
\newcommand{\RMLForgot}{0.69\xspace}
\newcommand{\RMLHHalf}{0.32\xspace}
\newcommand{\RMLHNinety}{0.31\xspace}
\newcommand{\RMLHalf}{0.33\xspace}
\newcommand{\RMLMonth}{0.44\xspace}
\newcommand{\RMLNew}{0.47\xspace}
\newcommand{\RMLNewMem}{0.39\xspace}
\newcommand{\RMLNoExp}{0.40\xspace}
\newcommand{\RMLNoReguess}{0.40\xspace}
\newcommand{\RMLNoReprompt}{0.38\xspace}
\newcommand{\RMLQuarter}{0.43\xspace}
\newcommand{\RMLRec}{0.21\xspace}
\newcommand{\RMSure}{0.59\xspace}
\newcommand{\RMUnfam}{0.57\xspace}
\newcommand{\RMUnsure}{0.35\xspace}
\newcommand{\RPartial}{0.45\xspace}
\newcommand{\RST}{0.59\xspace}
\newcommand{\RUnexpl}{0.44\xspace}
\newcommand{\RatingEnd}{6 May 2024\xspace}
\newcommand{\RecConcCIML}{3 to 25\xspace}
\newcommand{\RecConcCIST}{$-$7 to 15\xspace}
\newcommand{\RegGuess}{0.62\xspace}
\newcommand{\RegGuessSE}{0.07\xspace}
\newcommand{\RegMF}{0.65\xspace}
\newcommand{\RegMFSE}{0.09\xspace}
\newcommand{\RegML}{$-$0.21\xspace}
\newcommand{\RegMLSE}{0.10\xspace}
\newcommand{\SdGRec}{0.99\xspace}
\newcommand{\SdMLRec}{0.44\xspace}
\newcommand{\SplitDiffLevelCIST}{$-$0.02 to 0.21\xspace}
\newcommand{\SplitDiffLevelST}{0.09\xspace}
\newcommand{\SplitDiffMseCIST}{$-$0.05 to 0.28\xspace}
\newcommand{\SplitDiffMseST}{0.12\xspace}
\newcommand{\SplitDiffWithinCIST}{$-$0.08 to 0.12\xspace}
\newcommand{\SplitDiffWithinST}{0.02\xspace}
\newcommand{\SplitMembers}{77\xspace}
\newcommand{\SplitN}{330\xspace}
\newcommand{\SplitRG}{0.64\xspace}
\newcommand{\SplitRST}{0.59\xspace}
\newcommand{\StackGbmR}{0.56\xspace}
\newcommand{\StackOlsR}{0.59\xspace}
\newcommand{\StaleSE}{0.015\xspace}
\newcommand{\StaleSlope}{0.014\xspace}
\newcommand{\TransferShareSim}{74\xspace}
\newcommand{\ValN}{20,000\xspace}
\newcommand{\WRegConst}{0.32\xspace}
\newcommand{\WRegFamAbs}{0.30\xspace}
\newcommand{\WRegFamP}{0.080\xspace}
\newcommand{\WRegFamSE}{0.17\xspace}
\newcommand{\WRegM}{$-$0.29\xspace}
\newcommand{\WRegMP}{0.03\xspace}
\newcommand{\WRegMSE}{0.14\xspace}
\newcommand{\WRegNew}{$-$0.33\xspace}
\newcommand{\WRegNewSE}{0.31\xspace}
\newcommand{\WRegU}{$-$0.36\xspace}
\newcommand{\WRegUSE}{0.30\xspace}
\newcommand{\WatchCertHigh}{3.4\xspace}
\newcommand{\WatchCertLow}{0.6\xspace}
\newcommand{\WatchHigh}{10.2\xspace}
\newcommand{\WatchLow}{0.6\xspace}

\title{You Know What You Will Like:\\People Predict Their Own Ratings\\Better Than Recommender Systems}

\author{
  Zihan Chen \\
  Stevens Institute of Technology \\
  Hoboken, NJ, USA \\
  \texttt{zchen61@stevens.edu} \\
  \And
  Lei Nico Zheng \\
  University of Massachusetts Boston \\
  Boston, MA, USA \\
  \texttt{lei.zheng@umb.edu} \\
  \And
  Di Zhu \\
  Stevens Institute of Technology \\
  Hoboken, NJ, USA \\
  \texttt{dzhu1@stevens.edu} \\
}

\begin{document}
\maketitle

\begin{abstract}
Recommender systems usually learn what people want from what they have done. Platforms could also ask people directly, but asking is costly: \NoAnswerShare\% of the prompts in our data went unanswered. Theories of statistical prediction imply that a person's own forecast adds little to a recommender algorithm built from many people's behavior. Theories of dispersed knowledge imply that people hold information about their own taste that the data lack. We test the two with the MovieLens Beliefs dataset. In 2023--2024, a homepage prompt asked members how they expected to rate movies they said they had not seen. Of \NGuesses{} such expectations, \NLater{} from \NLaterUsers{} members were followed by a rating at least a day later. We compare each expectation with four benchmarks: the platform's own predicted rating, two collaborative-filtering models (matrix factorization and an item neighborhood), and a stack trained on other members' outcomes. The trained models use each member's latest ratings. People predict their own ratings more accurately than all four benchmarks (correlation \RG{}, against \RML{} for MovieLens and \RMF{} to \RST{} for the trained recommenders). Added to all the machine forecasts and features, the expectation lowers out-of-sample mean squared error by \PairGuessMSEShare\%. The gain holds when models are trained on earlier expectations and tested on later ones. A member's earlier expectations about other movies do not replace the current one. In ordering a member's own movies, the trained recommenders are statistically indistinguishable from the person. On old, heavily rated movies, the machines do as well or better. The results reconcile the two theories: the crowd's data suffice where they are rich, and the person adds information about the specific item. We derive four design principles for preference elicitation and recommender evaluation.
\end{abstract}


\section{Introduction}
\label{sec:intro}

Recommender systems decide much of what people watch, read and buy. At Netflix, the recommender ``influences choice for about 80\% of hours streamed'' \citep[p.~13:5]{gomezuribe2015}. To choose well, a platform has to know what each person will like, and it faces a design decision about how to learn it. It can infer preferences from what people and millions of others have done: the ratings they gave, the titles they finished, the items they clicked \citep{adomavicius2005,koren2009}. Or it can ask a person what they expect from a specific item before they choose it. Asking is costly. When MovieLens placed such a question on its homepage, \NoAnswerShare\% of the \NPrompts{} prompts went unanswered. Whether a question is worth that cost, which question to ask and about which items are decisions platforms make with little evidence. The stakes are highest where recommenders are weakest: among the movies MovieLens's own recommender ranked highest for a member, its predictions put the member's later ratings in the right order only \ConcMLRec\% of the time, where a coin flip scores 50\%.

The answer turns on who knows what, and two bodies of theory point in opposite directions. The first, which we call the crowd view, holds that statistical aggregation of many people's behavior beats individual judgment. Mechanical prediction matches or beats clinical judgment \citep{grove2000,kleinberg2018}, computers judge people's personalities more accurately than their friends and spouses do \citep{youyou2015}, and recommenders predict which jokes a person will find funny better than the people closest to them \citep{yeomans2019}. People are also poor forecasters of their own enjoyment \citep{kahneman1992,wilson2005}; in one well-known study, they predicted their reaction to an event more accurately from a stranger's report of it than from information about the event itself \citep{gilbert2009}, which is exactly the logic of collaborative filtering. On this view, what a member could tell the platform is already in its data, and asking is redundant.

The second, the view of dispersed knowledge, holds that much of the knowledge relevant to a decision is knowledge of particular circumstances, dispersed among individuals and beyond the reach of any central authority \citep{hayek1945}. Research on human--AI collaboration builds on the same idea: people hold ``unique human knowledge'' that an algorithm lacks \citep{fugener2021}, and when they do, it pays to ask them for it \citep{ibrahim2021}. A person forecasting their own enjoyment of a specific movie is a clear case. They may know that they love the director's earlier work but never rated it, that the trailer left them cold, or that a friend whose taste they trust could not stop talking about it. On this view, the person holds information about the match between person and movie that no rating records, and asking adds to what the platform knows.

The two views make opposite predictions about the value of asking, and neither settles it for recommender design. The evidence for the crowd view comes from people forecasting \emph{someone else}: another member's ratings \citep{krishnan2008}, a person's taste in jokes \citep{yeomans2019}. The evidence for complementarity comes mostly from experts and managers inside organizations \citep{blattberg1990,ibrahim2021,fugener2021}. Information systems research on recommendation agents has asked how their design, including how they elicit preferences, shapes consumers' decisions and their evaluations of the agent \citep{xiao2007}, and how displayed predictions shape ratings \citep{adomavicius2013}; we know of no study in this line of work that measures how much a consumer's own forecast adds to the agent's. Nor does either view say when it holds. The MovieLens Beliefs dataset \citep{aridor2024} makes the test possible. Its developers show that members' expectations are on average unbiased and that recommendations shift them \citep{aridor2022informational}, but they do not set those expectations against the recommender or ask what they add to it (Table~\ref{tab:position}). We do.

We ask three questions.
\begin{description}
  \item[RQ1.] How accurately do people forecast their own rating of a movie they have not seen, compared with the platform's recommender and with strong recommenders built from the same rating data?
  \item[RQ2.] Where does the person's advantage come from: does it lie beyond the member's and the movie's average ratings, and is it larger where the machine's information is thin?
  \item[RQ3.] What does the guess add to what a recommender already knows, can summaries of the member's earlier guesses stand in for it, and does the member's certainty add to it?
\end{description}

From March 2023 to May 2024, MovieLens asked members on its homepage how they thought they would rate movies they had not seen. Of \NGuesses{} such guesses, \NLater{} were followed by a rating the member entered at least a day later, about a movie with no earlier rating or reported viewing; these \NLater{} member--movie pairs from \NLaterUsers{} members are our main sample. On them we compare each guess with the prediction of MovieLens's own recommender and with recommenders we train on up to \NTrainPool{} ratings entered before the guess, updated with everything the member had rated up to that moment: matrix factorization, an item-neighborhood model, and a stack that combines every machine forecast with content and history features and is trained on other members' outcomes in the study. All forecasts are scored on the same pairs, so differences between them do not come from different samples. The headline result is simple. The member's guess correlates \RG{} with the later rating; the MovieLens prediction correlates \RML{}, and the trained recommenders between \RMF{} and \RST{}. The guess also predicts the part of the rating that the member's and the movie's averages do not explain, \RBaseG{} against at most \RBaseMachMax{} for the machines, and it misses the later rating by \MAEG{} stars on average, against at least \MAEMachMin{} for any machine. Added to all the machine forecasts and features, the guess lowers the out-of-sample mean squared error by \PairGuessMSEShare\% (from $R^2 = \AblMachRsq$ to \AblGuessRsq), and the gain holds when models are trained on earlier guesses and tested on later ones. Summaries of the member's earlier guesses about other movies, including summaries weighted by each movie's similarity to the target, add no detectable value and do not replace the current guess. Ordering one member's own movies is where the crowd's data catch up. Given two movies a member had not yet seen and later rated differently, the member's guesses put them in the right order \ConcG\% of the time, MovieLens \ConcML\% and the trained recommenders \ConcMF\% to \ConcMachMax\%. The person clearly beats MovieLens, and the trained recommenders are statistically indistinguishable from the person.

The comparison holds within fixed follow-up windows, for movies never prompted again before the rating, for movies absent from the member's recorded recommendation lists, and after reweighting to the guesses about movies with a realistic chance of being rated. Guesses about movies members had rated years earlier and then reported as unseen correlate \RGForgot{} with a rating the guess could not have influenced. The results also mark where the person's edge ends. On those old, heavily rated movies the machines do as well as the person or better, and for ratings entered months after the guess the trained recommenders order movies as well as the person or better, while the person's lead in correlation persists. Together the results reconcile the two views. The crowd view holds where the platform's data on a member and a movie are rich, and in ordering a member's own movies, where the trained recommenders are statistically indistinguishable from the person. The dispersed-knowledge view holds for how much a member will like the specific movie asked about, where the person adds information that the rating data lack.

The paper makes contributions from both theoretical and practical perspectives. On the theoretical side, first, we provide, to our knowledge, the first field comparison of people's forecasts of their own enjoyment of specific unseen items with a production recommender and with trained recommenders from the two main families of collaborative filtering, on the same items and with later outcomes. When the target is one's own rating of a specific movie, the person is the more accurate forecaster, and the gap is large. Second, we reconcile the crowd view and the view of dispersed knowledge by tracing when each holds. The person's guess carries information that the machine forecasts and features lack: beyond the member's and the movie's averages, beyond every machine forecast, and beyond simple and similarity-weighted summaries of the member's earlier guesses, while the machines match the person where their data are rich. In recommendation the platform is usually the better-informed party; for the rating of a specific item, the user knows something the platform does not. Third, on the methodological side, we develop a set of design checks for comparing self-forecasts with machines in field data: aligning the clocks of the event logs, separating corrections and known earlier experience from new viewings, fixing follow-up windows, using items rated before the forecast, whose outcome the forecast cannot have shaped, and reweighting for selection into rating on a target where the weights balance. The most consequential check is what the benchmark is allowed to know: monthly snapshots that missed members' latest ratings understated the trained recommenders and overstated the person's lead in ordering.

In terms of practical contributions, we turn the findings into four design principles for preference elicitation and recommender evaluation (Table~\ref{tab:principles}). First, platforms should ask for the expected rating of a specific item: a single guess predicts the later rating better than all the machine signals combined and adds to all of them, while the member's certainty adds little. Second, platforms should ask about the items being decided on, because summaries of the member's earlier guesses do not replace the current one. Third, recommenders should be evaluated on the candidates they re-rank. Among the movies the recommender ranked in a member's top hundred, its predictions order the member's later ratings correctly \ConcMLRec\% of the time, against \ConcGRec\% for the member's own guesses, a weakness that accuracy measured across all items hides. Fourth, the benchmark should know what the platform knows, so a member's latest ratings should be folded in before a recommender is judged against people.

The rest of the paper proceeds as follows. Section~\ref{sec:lit} sets out the theoretical background and Section~\ref{sec:theory} derives the competing predictions. Section~\ref{sec:data} describes the data and the machines. Section~\ref{sec:main} reports the main comparison, Section~\ref{sec:real} tests its robustness, Section~\ref{sec:where} asks where the person's edge is larger, and Section~\ref{sec:value} asks what the guess adds to the machines. Section~\ref{sec:discussion} discusses theoretical and practical implications and limitations.

\section{Theoretical Background}
\label{sec:lit}

Whether a platform should ask people what they expect depends on who holds the information that predicts their experience. Two bodies of theory answer differently: the crowd view, built on statistical prediction and on how people forecast their own enjoyment, and the view of dispersed knowledge, built on local knowledge and on human--AI complementarity. A third literature, on recommender systems in information systems research, defines the design problem and the methods the comparison needs. Table~\ref{tab:position} sets out the closest empirical studies.

\subsection{The crowd view: statistical prediction and the limits of self-forecasts}

A long tradition finds that statistical rules match or beat expert judgment. In a meta-analysis of clinical and mechanical prediction, \citet{grove2000} find that mechanical prediction is typically at least as accurate as clinical judgment. Machine learning has extended the comparison to high-stakes decisions such as bail, where predictions built from administrative data can outperform judges \citep{kleinberg2018}, and to judgments about people themselves: models built from Facebook Likes judge personality more accurately than friends, family and spouses do \citep{youyou2015}. Recommendation has been a central case. On MovieLens, collaborative filtering predicted members' ratings better on average than active members who worked from the same profiles \citep{krishnan2008}, and recommender systems predicted which jokes a person would find funny better than strangers, friends or family \citep{yeomans2019}. A related literature studies whether people trust algorithms, finding both aversion after seeing an algorithm err \citep{dietvorst2015} and a preference for algorithmic over human advice \citep{logg2019}.

Psychology adds reasons to doubt self-forecasts in particular. People mispredict how their liking will change with repeated exposure \citep{kahneman1992}, and they overestimate how strongly and how long they will react to future events \citep{wilson2005}. Most strikingly for recommendation, \citet{gilbert2009} found that people predicted their enjoyment of a future event more accurately when they knew only how another person had reacted to it than when they knew about the event itself, while believing the opposite. Collaborative filtering is this ``surrogation'' at scale: it predicts a member's reaction from the reactions of similar members. On this view, a recommender built from many people's ratings should beat the member's own forecast, and the forecast should add little to it.

\subsection{The view of dispersed knowledge: local knowledge and human--AI complementarity}

\citet{hayek1945} argued that much of the knowledge on which economic decisions depend is dispersed among individuals as knowledge of particular circumstances, and that no central authority can collect it in full. Recommender systems are central in this sense: they aggregate the recorded behavior of many people, but a person's reaction to a trailer, their memory of a director's earlier work or a friend's enthusiasm for a movie reaches the platform only if the person reports it.

Research on human--AI collaboration applies the same idea to prediction. Combining a model with a manager's forecast beats either alone \citep{blattberg1990}. \citet{ibrahim2021} show that when people hold information an algorithm lacks, it can pay to ask them for that information directly rather than for a full forecast. \citet{fugener2021} show that advice from an AI can erode the ``unique human knowledge'' that makes a group of people valuable, and \citet{fugener2022} find that people are poor judges of their own metaknowledge and therefore delegate poorly to an AI. On this view, a person forecasting their own enjoyment of a specific movie holds information about the match that no rating records, and the forecast should add to what the machine knows.

The evidence behind the two views comes from different forecasters. In the human--machine comparisons above, the human predicts something about \emph{someone else}: a defendant, a patient, a friend's personality, another member's ratings. In the complementarity studies, the human is an expert or a manager inside an organization. A consumer forecasting their own experience of an item is a different case, and the one in which a human is most likely to hold information that the data do not record.

\begin{table}[t]
\centering
\caption{The closest studies and this paper.}\label{tab:position}
\footnotesize
\begin{tabularx}{\linewidth}{>{\raggedright\arraybackslash}p{2.6cm}>{\raggedright\arraybackslash}X>{\raggedright\arraybackslash}p{2.0cm}>{\raggedright\arraybackslash}X}
\toprule
Study & Who forecasts what & Compared with a recommender & What the study establishes \\
\midrule
\citet{krishnan2008} & Active MovieLens members predict another member's ratings from that member's profile & Yes & Collaborative filtering is more accurate on average; some people beat it \\
\citet{yeomans2019} & Strangers, friends and family predict which jokes a person will find funny & Yes & The recommender is more accurate; people are averse to relying on it \\
\citet{aridor2024} & MovieLens members state expected ratings and certainty for movies they have not seen & No & Releases the belief data and documents the prompt \\
\citet{aridor2022informational,aridor2023} & The same kind of beliefs, elicited in a field experiment & No & Beliefs are unbiased on average and less accurate when members are less sure; recommendations shift beliefs and consumption \\
This paper & Members forecast their own later rating of a specific unseen movie & Yes: the production system and three trained recommenders & The member's guess is the more accurate rating forecast and adds to all machine information; in ordering, the trained recommenders are statistically indistinguishable from the person \\
\bottomrule
\end{tabularx}
\end{table}

\citet{aridor2022informational,aridor2023} come closest. In a field experiment on MovieLens they elicit members' beliefs about unseen movies and show that recommendations shift those beliefs and, through them, what people watch. In validating the beliefs, they report that beliefs are on average unbiased forecasts of later ratings, that the gap between expected and later rating grows with the member's stated uncertainty, and that members are surer about extreme expectations \citep[Appendix~A]{aridor2022informational}. Their focus is how recommendations change beliefs and consumption. They do not compare the accuracy of beliefs with the recommender's predictions or ask what beliefs add to them, which are our questions, and the MovieLens Beliefs dataset \citep{aridor2024} lets us ask them across the whole platform.

\subsection{Recommender systems in information systems research}

Recommender systems infer preferences from behavior and from ratings of items already experienced \citep{adomavicius2005,koren2009}. They ask directly mainly at the start, when a new member rates a few chosen items \citep{rashid2002}, and through active learning, which picks the items whose ratings would teach the model most \citep{elahi2016}. Conversational recommenders ask about attributes and preferences \citep{christakopoulou2016}, and \citet{ekstrand2016} argue that behavioral data alone are not enough and that recommenders should listen to what users say. These designs ask people to report on items they have already experienced, or on general tastes. The MovieLens prompt studied here asks for something different: a forecast of one's own reaction to a specific item not yet seen, which is the forecast a recommender exists to make.

Information systems research has studied recommendation agents mainly as decision aids. \citet{xiao2007} develop a model of how an agent's use and design, including how it elicits consumers' preferences, shapes consumers' decision processes and outcomes and their evaluations of the agent, and \citet{adomavicius2013} show that a displayed predicted rating anchors the rating a consumer then gives \citep[see also][]{cosley2003}. In this work the consumer is mainly the one who receives the agent's output. We measure the information that flows the other way: how much a consumer's own forecast adds to the agent's prediction.

Two further strands bear on how the comparison has to be made. First, ratings are missing not at random: people rate what they chose to watch, and they choose what they expect to like \citep{marlin2009,steck2010}. The same selective-labels problem complicates human--machine comparisons elsewhere \citep{lakkaraju2017,kleinberg2018}. Second, ratings are noisy and malleable: the same user rates the same item differently on different occasions \citep{amatriain2009}, and a displayed prediction pulls ratings toward itself \citep{cosley2003,adomavicius2013}. We examine how selection into rating bears on the comparison, use movies members had rated before reporting them unseen to test whether the guess predicts a rating it could not have influenced, and report how consistently members rate the same movie twice.

\section{Competing Predictions}
\label{sec:theory}

\subsection{A simple account of a rating}

Write member $i$'s later rating of movie $m$ as
\begin{equation}
  y_{im} \;=\; q_m \;+\; \mu_i \;+\; \tau_{im} \;+\; \varepsilon_{im}.
  \label{eq:rating}
\end{equation}
Here $q_m$ is the movie's common appeal, the part every member's rating shares; $\mu_i$ is how generously member $i$ uses the scale; $\tau_{im}$ is the personal match between this member and this movie; and $\varepsilon_{im}$ is the noise of the occasion: the mood, the company and the evening on which the movie was watched, and the imprecision of turning an experience into stars. No forecast made before viewing can anticipate $\varepsilon_{im}$.

The two forecasters see different parts of this equation.

\paragraph{The machine.} A recommender sees the ratings of hundreds of thousands of members. It estimates $q_m$ precisely for a movie many people have rated and $\mu_i$ precisely for a member with a long history. It estimates $\tau_{im}$ too, and collaborative filtering is designed to do so: a latent-factor model gives each member and each movie a vector of tastes and predicts their match from the inner product of the two vectors \citep{koren2009}, and a neighborhood model predicts it from the member's ratings of similar movies \citep{sarwar2001,koren2010}. Either way, the machine learns the part of the match that has already shown up in the member's ratings. Whatever part of the match the ratings have not revealed is invisible to it.

\paragraph{The person.} The member sees much of what signals $q_m$: the movie's reputation, its reviews and word of mouth. What the member also sees is their own reaction to what they know about the movie, such as its premise, its cast, its trailer or a friend's recommendation, and their sense of their own tastes, including tastes they have never expressed in a rating. These are signals about $\tau_{im}$ that the rating data do not record. The member's forecast is also imperfect: people misjudge how much they will enjoy things \citep{kahneman1992,wilson2005}, and a guess turns a vague expectation into stars.

\subsection{Two questions, two views}

Two questions are easy to run together. The first is which forecaster is more accurate on its own. The second is whether the person's forecast improves a forecast that already uses the machine's information. A machine can be the better forecaster and still gain from the person's input, if the person's errors are not the machine's errors. The views of Section~\ref{sec:lit} differ less on the first question than on the second, because they disagree about where the person's information comes from.

On the crowd view, what a member knows about their taste for a movie is mostly what their past ratings and the reactions of similar members already reveal. The member's forecast may be accurate, but it repeats what the data show: it should add little to a recommender that uses those data, and any systematic way in which a member's expectations differ from the machine's should be visible in the member's earlier guesses. On the view of dispersed knowledge, the member holds signals about the match that no rating records, and the match is what a recommender is for: once a movie's common appeal and the member's generosity are known, what remains is whether \emph{this} member will like \emph{this} movie. A guess should then carry information that the machine forecasts lack, and that information should concern the movie at hand.

\subsection{Predictions}

Table~\ref{tab:predictions} sets out what each view predicts for each question and where the paper tests it. Four of the predictions separate the views directly.

\begin{enumerate}
  \item \emph{Beyond the averages.} If the person's information concerns the match, the guess should predict the part of the later rating that the member's and the movie's average ratings do not explain, and predict it better than the machines.
  \item \emph{Beyond the machine.} Added to what a recommender knows before asking, the guess should improve a held-out forecast of the later rating.
  \item \emph{Not replaced by earlier guesses.} If the information concerns this member and this movie, summaries of the member's earlier guesses about other movies should not substitute for the guess about the movie at hand.
  \item \emph{Where the machine knows least.} The person's edge should be larger where the machine's evidence about the match is thin: for movies outside the genres a member usually rates, for members with short histories, and for movies few people have rated. New releases are a mixed case: the machine knows little about them, but so does the person, who has only trailers and reviews to go on.
\end{enumerate}
A last prediction concerns the person's own uncertainty. If members know which of their forecasts rest on real information, their stated certainty should tell a platform when to trust the guess over the machine, beyond the simple fact that people are surer about extreme expectations.

\begin{table}[t]
\centering
\caption{What each view predicts.}\label{tab:predictions}
\footnotesize
\begin{tabularx}{\linewidth}{>{\raggedright\arraybackslash}p{3.0cm}>{\raggedright\arraybackslash}X>{\raggedright\arraybackslash}X>{\raggedright\arraybackslash}p{2.0cm}}
\toprule
Question & Crowd view & Dispersed-knowledge view & Tested in \\
\midrule
Which forecaster is more accurate? & The machine, or no clear winner & The person, for the rating of the movie at hand & Section~\ref{sec:main} \\
Beyond the averages & The guess carries little beyond the member's and the movie's averages & The guess predicts the part of the rating the averages leave unexplained & Section~\ref{sec:main} \\
Beyond the machine & Little held-out gain from adding the guess & A substantial held-out gain & Section~\ref{sec:value} \\
Earlier guesses & Summaries of earlier guesses capture what the member adds & The current guess adds beyond them & Section~\ref{sec:value} \\
Where the edge is larger & No pattern, or none to explain & Larger where the machine's data are thin; the machines catch up where their data are rich & Sections~\ref{sec:real}--\ref{sec:where} \\
Certainty & Uninformative beyond the guess & Marks the guesses to trust over the machine & Sections~\ref{sec:where}--\ref{sec:value} \\
\bottomrule
\end{tabularx}
\end{table}

\section{Setting and Data}
\label{sec:data}

\subsection{MovieLens and the belief prompt}

MovieLens is a non-commercial movie recommender run by the GroupLens research lab at the University of Minnesota \citep{harper2015}. Members rate movies they have seen on a scale from half a star to five stars, and the platform's collaborative-filtering recommender turns those ratings into a personalized predicted rating for movies a member has not rated. The predictions are displayed next to movies and drive the ``top picks'' row on the homepage. MovieLens does not stream movies, so a rating is the member's own record of a movie watched elsewhere, and the platform does not observe when the movie was watched.

From \PromptStart{} to \PromptEnd{}, the second row of the MovieLens homepage carried a belief prompt \citep{aridor2024}. Each prompt showed a movie title and asked whether the member had seen it. A member who answered ``not seen'' was asked how they thought they would rate the movie, on the same half-star scale as ordinary ratings, and how sure they were, on a five-point scale. Movies entered the prompt from three sources: a broad sample of popular, highly rated, recent, trending and randomly chosen titles; a ``possible recommendation'' sample drawn at random from the hundred movies the recommender ranked highest for the member among that month's eligible movies the member had not rated; and one recent release \citep{aridor2024}. In a companion field experiment using a similar survey, the developers report that members saw only the title and poster and could not see the platform's predicted rating while answering \citep{aridor2022informational}.

The release records every prompt shown, every answer, and, for each prompt, the recommender's predicted rating for that member and movie. It also includes the complete rating histories of the members who answered and of other members who were shown prompts. We refer to the member's answer as the \emph{guess}, to the platform's predicted rating as the \emph{MovieLens prediction}, and to a rating the member entered after the guess as the \emph{later rating}.

\subsection{Sample}

Table~\ref{tab:funnel} traces the sample. The prompt was shown \NPrompts{} times to \NPromptUsers{} members. Members answered ``not seen'' and gave a guess for \NGuesses{} distinct member--movie pairs, from \NGuessUsers{} members about \NGuessMovies{} movies; when the same movie was asked about twice, we keep the first answer. Ratings in the release run to \RatingEnd{}, so a guess could be followed for between \FollowMin{} and \FollowMax{} days.

\begin{table}[t]
\centering
\caption{Sample.}\label{tab:funnel}
\small
\begin{tabular}{lrr}
\toprule
 & Count & Members \\
\midrule
Prompts shown & 3,004,084 & 22,382 \\
\quad no answer & 2,967,057 &  \\
\quad answered ``seen'' & 8,565 &  \\
\quad answered ``not seen'' with a guess & 28,462 &  \\
Distinct member--movie guesses (first answer kept) & 26,020 & 3,200 \\
\quad first rating within a day of the guess (a correction) & 284 &  \\
\quad first rating one day or more after the guess & 570 & 285 \\
\qquad movie rated before the guess (excluded) & 12 &  \\
\qquad viewing reported before the guess (excluded) & 7 &  \\
\qquad main sample & 551 & 277 \\
\quad movie rated 30 days or more before the guess (previously rated) & 238 & 108 \\
Movies rated again in a prompt, 30 days or more after the first rating & 1,444 & 615 \\
Ratings used to train the machines & 32,993,711 &  \\
\bottomrule
\end{tabular}
\par\vspace{2pt}\parbox{\linewidth}{\footnotesize\emph{Notes:} Prompts ran from 13 March 2023 to 1 May 2024. Training ratings combine the rating histories in the release with the public MovieLens 32M ratings. Each model is trained on ratings entered before the month of the guess and folds in the member's ratings entered before the guess.}
\end{table}

\paragraph{Later ratings.} For each guess we take the first rating the member entered after it. The guess is \emph{followed by a later rating} when that first rating came at least one day after the guess. The one-day gap matters. Of the \NQuick{} first ratings entered within a day, \QuickHalfHour{} came within thirty minutes of the member answering ``not seen'', too soon to have watched a feature film; these look like members correcting their answer rather than reporting a new viewing, and those guesses have no later rating. We also drop guesses about movies the member is known to have seen before: \NExclPriorRating{} that the member had already rated before the guess, and \NExclPriorView{} that the member later marked as seen in another prompt with a viewing date before the guess. The main sample has \NLater{} member--movie pairs from \NLaterUsers{} members and \NLaterMovies{} movies, with later ratings entered a median of \LagMedian{} days after the guess. All comparisons below are made on these pairs, the guesses that were followed by a recorded rating. Section~\ref{sec:real} examines how that selection, the timing of ratings and the sample definition bear on the comparison.

\paragraph{Previously rated movies.} For \NForgot{} guesses (\NForgotUsers{} members), the member had already rated the movie at least thirty days before answering ``not seen''. These guesses can be compared with a rating that was entered before the guess and therefore cannot have been influenced by it.

\paragraph{Repeat ratings.} When a prompt showed a movie the member had already rated, the member could mark it as seen and rate it again. For \NRetest{} such pairs (\NRetestUsers{} members) the earlier rating was at least thirty days old. The agreement between the two ratings shows how consistently people rate the same movie twice.

\paragraph{Timestamps.} The rating histories record US Central time, while the prompt log records Coordinated Universal Time. The same rating appears in both files exactly five hours apart in summer and six in winter; we convert the histories to Coordinated Universal Time before ordering any events. Without this step, a rating entered within a few hours after a guess would appear to precede it.

\subsection{The machines}

We compare the guess with the MovieLens prediction and with recommenders we build ourselves, so that the comparison does not rest on one production system. The release describes the MovieLens prediction as coming from the platform's standard collaborative-filtering algorithm \citep{aridor2024}, and we treat it as the platform's forecast. Our own recommenders are trained on ratings entered before the first day of the month in which the guess was made, and every member--movie pair with a guess is removed from training, so no model has seen the member's own rating of a movie they guessed about. Members keep rating after the cutoff, and many members' first recorded rating came during the month: by the time of their guess, such members had typically entered about a hundred ratings (median \NewMemHistMedian), none of them before the cutoff. Each forecast therefore folds in the member's own ratings entered before the guess: the movie side of the model stays as trained at the cutoff, and the member's effects are re-estimated from everything the member had rated before the moment of the guess. The training data combine the rating histories in the release with the public MovieLens 32M dataset \citep{harper2015}, after dropping the 32M accounts that duplicate members of the release (Appendix~\ref{app:machines}). Each monthly model has between \NTrainMin{} and \NTrainMax{} ratings in force at its cutoff, \NTrainPool{} distinct ratings in all. The public 32M data end in October 2023, so later monthly models add only the ratings of members in the release.

\begin{itemize}
  \item \emph{Matrix factorization}: a biased latent-factor model \citep{koren2009} with \MFk{} factors, fitted by stochastic gradient descent.
  \item \emph{Item neighborhood}: item-based collaborative filtering \citep{sarwar2001}. The forecast adds to a user + movie baseline a weighted average of the member's own deviations from that baseline on the \KnnK{} rated movies most similar to the target, with similarities shrunk toward zero when few members rated both movies \citep{koren2010}.
  \item \emph{Stacked machine}: a linear regression of the later rating on every machine forecast and on what a platform knows about the movie and the member before the guess (the rating counts of each, the movie's age and genres, the member's familiarity with those genres, the prompt source and the member's recommendation lists), trained on other members' outcomes in the study and evaluated on held-out members.
  \item \emph{User + movie baseline}, \emph{movie mean} and \emph{user mean}: the overall mean plus a member effect and a movie effect; the movie's average rating among all members; and the member's own average rating.
\end{itemize}

Matrix factorization and the neighborhood model were tuned on a forecasting task of the same shape: ratings that study members entered in March and April 2023, predicted from earlier data. The stack has an advantage the historical recommenders lack: it is trained on outcomes from the same study, from other members. Section~\ref{sec:value} also fits models on earlier guesses only and tests them on later ones. The movie mean carries only the crowd's view of the movie and nothing personal.

\subsection{Measures of accuracy}

Two measures carry the comparison. The first is the correlation between a predictor and the later rating across all \NLater{} pairs, which uses every member. The second is \emph{pairwise ordering accuracy}, which matches what a recommender does when it ranks movies for one member. Take two movies the same member later rated differently, and ask whether a predictor put them in the right order; ties in the predictor count as half right. We compute the share of correctly ordered pairs for each member who has at least one such pair and average across members, so that each member counts equally. The measure equals one half for a coin flip and one for perfect ordering, and it is unaffected by how each person uses the rating scale. It rests on the \ConcMembers{} members who later rated at least two movies differently.

We also report the correlation within member, using deviations from each member's own mean; the correlation \emph{beyond the baseline}, between the parts of the later rating and of the forecast that the user + movie baseline, estimated from data before the guess, does not explain; the mean absolute error; and the mean absolute error after a least-squares linear rescaling of each predictor, fitted with each member's movies held out together. Confidence intervals come from a bootstrap that resamples members, \Boot{} draws, so that all of a member's movies stay together and a member drawn twice counts twice.

\section{People Forecast Their Own Ratings Better Than the Machines}
\label{sec:main}

Table~\ref{tab:main} compares the predictors on the \NLater{} member--movie pairs in the main sample. The member's own guess is the most accurate forecast of the later rating, against the production recommender, the two recommenders we trained, and the stacked machine. Its advantage in correlation and error is clear against every machine. In ordering one member's own movies, it clearly beats MovieLens, and the trained recommenders are statistically indistinguishable from it.

\begin{table}[t]
\centering
\caption{How well each predictor anticipates the later rating.}\label{tab:main}
\small
\setlength{\tabcolsep}{4pt}
\begin{tabular}{lrcrrrrr}
\toprule
 & \multicolumn{2}{c}{Ordering accuracy (\%)} & \multicolumn{3}{c}{Correlation with later rating} & \multicolumn{2}{c}{Mean abs.\ error} \\
\cmidrule(lr){2-3}\cmidrule(lr){4-6}\cmidrule(lr){7-8}
Predictor & Share & 95\% CI & All & Within member & Beyond baseline & Raw & Rescaled \\
\midrule
Person's own guess & 71.6 & [64.9, 77.9] & 0.69 & 0.50 & 0.63 & 0.61 & 0.64 \\
\addlinespace
Stacked machine & 66.2 & [57.6, 74.2] & 0.59 & 0.48 & 0.47 & 0.72 & 0.72 \\
Matrix factorization & 63.8 & [55.1, 72.2] & 0.54 & 0.35 & 0.39 & 0.73 & 0.74 \\
Item neighborhood & 66.9 & [58.0, 75.2] & 0.56 & 0.47 & 0.41 & 0.73 & 0.73 \\
MovieLens prediction & 56.1 & [46.6, 65.0] & 0.34 & 0.09 & 0.11 & 0.84 & 0.84 \\
User + movie baseline & 57.5 & [48.9, 65.7] & 0.41 & 0.20 & -- & 0.79 & 0.81 \\
Movie mean & 61.5 & [53.2, 70.3] & 0.25 & 0.20 & 0.01 & 0.86 & 0.89 \\
User mean & 46.5 & [38.7, 54.4] & 0.31 & 0.02 & 0.00 & 0.85 & 0.85 \\
\bottomrule
\end{tabular}
\par\vspace{2pt}\parbox{\linewidth}{\footnotesize\emph{Notes:} 551 member--movie pairs (389 movies) from 277 members. Ordering accuracy: for each of the 77 members with at least one pair of later-rated movies whose ratings differ (1,399 pairs), the share of pairs the predictor puts in the right order (ties count one half), averaged over members. ``Within member'' correlates deviations from each member's own mean (47 members with three or more rated movies). ``Beyond baseline'' correlates the later rating minus the user + movie baseline with the prediction minus the same baseline. ``Rescaled'' applies a least-squares linear rescaling of each predictor, fitted with each member's movies held out together. The stacked machine is a linear regression trained on other members' later ratings in the study and evaluated on held-out members. Confidence intervals resample members (2,000 draws).}
\end{table}

\paragraph{Correlation and error.} The guess correlates \RG{} with the later rating. The trained recommenders reach \RMF{} (matrix factorization), \RKN{} (item neighborhood) and \RST{} (stacked machine), and the MovieLens prediction \RML. The difference from the strongest of them, the stacked machine, is \DiffRST{} (95\% interval \DiffRCIST); from matrix factorization \DiffRMF{} (\DiffRCIMF); and from MovieLens \DiffRML{} (\DiffRCIML). The guess is also nearly unbiased: on average the later rating differs from it by \BiasG{} stars, while the MovieLens prediction overstates the later rating by \BiasAbsML{} stars, the item neighborhood by \BiasAbsKN{} and matrix factorization by \BiasAbsMF{}. The guess misses the later rating by \MAEG{} stars on average, the most accurate machine by \MAEMachMin{}; after each predictor is rescaled by a linear fit estimated out of sample, by \MAEcalG{} and \MAEcalMachMin{} stars.

\paragraph{Beyond the averages.} The theory in Section~\ref{sec:theory} says the person's information concerns the match between member and movie, beyond the movie's common appeal and the member's generosity. Subtracting the user + movie baseline, estimated from ratings entered before the guess, from both the later rating and each forecast leaves the part of each that the two averages do not explain. The guess correlates \RBaseG{} with that part of the later rating; the machines at most \RBaseMachMax{} (difference from the stacked machine \DiffRBaseST, 95\% interval \DiffRBaseCIST). The person's advantage therefore does not come only from knowing which movies are good or how generously they rate; it extends to the part of the rating that the baseline leaves to the particular member and movie.

\begin{figure}[t]
  \centering
  \includegraphics[width=0.78\linewidth]{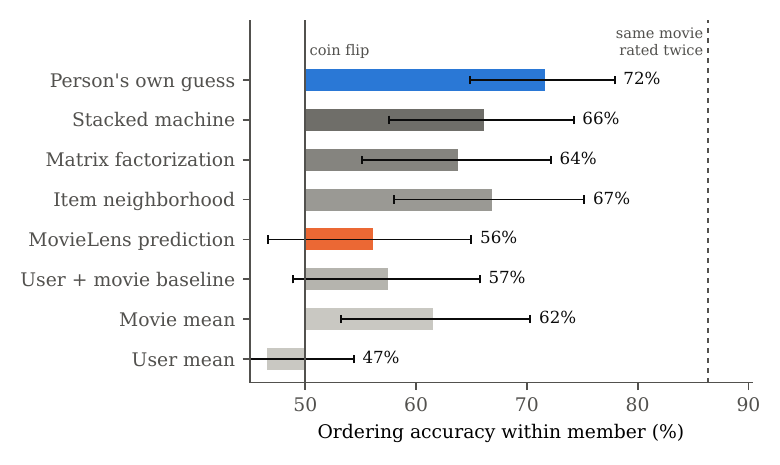}
  \caption{Ordering accuracy within member: the share of a member's pairs of differently rated movies that each predictor puts in the right order, averaged over the \ConcMembers{} members with at least one such pair. Bars show 95\% intervals from resampling members. The dashed line is a reference, not a ceiling: how often two ratings of the same movie by the same member, entered a median of \CeilGapYears{} years apart, order the member's movies the same way.}
  \label{fig:order}
\end{figure}

\paragraph{Ordering.} Among the \ConcMembers{} members who later rated at least two movies differently, the guess puts two movies in the right order \ConcG\% of the time (95\% interval \ConcCIG\%). The MovieLens prediction does so \ConcML\% of the time, against 50\% for a coin flip, and the person's advantage over it is \DiffConcML{} percentage points (\DiffConcCIML). The trained recommenders come closer: matrix factorization orders \ConcMF\%, the item neighborhood \ConcKN\% and the stacked machine \ConcST\%. Their differences from the person, \DiffConcKN{} to \DiffConcMF{} points, have intervals that include zero: in ordering one member's movies, the trained recommenders and the person are statistically indistinguishable (Appendix~\ref{app:robust} reports the other measures on these members). The movie's average rating among all members, which knows nothing about the person, orders \ConcIM\% of pairs correctly, more than the MovieLens prediction. Figure~\ref{fig:order} shows the comparison.

\begin{figure}[t]
  \centering
  \includegraphics[width=0.78\linewidth]{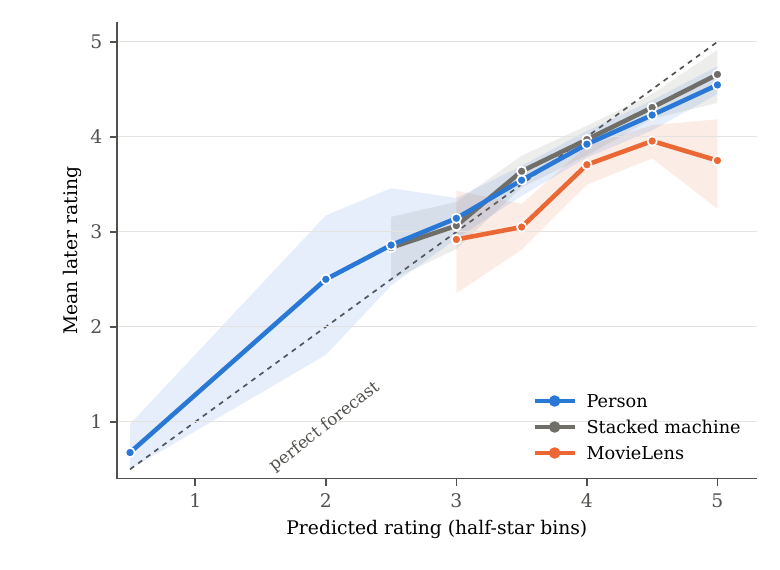}
  \caption{Mean later rating by predicted rating, in half-star bins with at least 15 pairs. Shaded bands are 95\% intervals from resampling members. The dashed line marks a perfect forecast.}
  \label{fig:binned}
\end{figure}

\paragraph{Calibration.} Figure~\ref{fig:binned} plots the average later rating against the guess, the stacked machine and the MovieLens prediction. The guess is calibrated across the whole scale: movies guessed at half a star average \CalGHalf{} stars later, movies guessed at three stars \CalGThree, and movies guessed at five \CalGFive. The stacked machine is calibrated over the narrower range its forecasts cover, as a forecast of a conditional mean should be; what separates it from the guess is held-out loss, not the spread of the forecasts. The MovieLens prediction is not calibrated at the top of the scale: movies it predicted at five stars average \CalMLFive{} later, lower than those it predicted at four and a half (\CalMLFourHalf).

\paragraph{The guess, MovieLens and matrix factorization in one regression.} Regressing the later rating on the guess and the MovieLens and matrix-factorization predictions, with errors clustered by member, the guess carries a coefficient of \RegGuess{} (standard error \RegGuessSE) and matrix factorization \RegMF{} (\RegMFSE): each holds information the other lacks. The MovieLens prediction takes a small negative weight once the other two are known (\RegML, standard error \RegMLSE). Section~\ref{sec:value} turns this into an out-of-sample question about what the guess adds to all the machine information.

\paragraph{By source of the prompt.} Table~\ref{tab:source} splits the comparison by how the movie entered the prompt. Most pairs in the main sample (\NRec) came from the ``possible recommendation'' source, drawn from the recommender's top hundred for the member. Among these, MovieLens's predictions are tightly bunched (standard deviation \SdMLRec{} stars among the rated movies, against \SdGRec{} for the guesses), and its correlation with the later rating is \RMLRec, against \RGRec{} for the guess and \RMFRec{} for matrix factorization. Re-ranking its own top candidates is a central task of a recommender, and here the production system does it poorly: among these movies the guess orders a member's later ratings correctly \ConcGRec\% of the time and MovieLens \ConcMLRec\% (interval for the difference \RecConcCIML{} points). The trained recommenders do better than MovieLens: the stacked machine orders \ConcSTRec\%, statistically indistinguishable from the person (95\% interval for the difference \RecConcCIST{} points). In the broad sample of movies, where the recommender's predictions spread more widely, the guess still leads (\RGBroad{} against \RMLBroad), as it does for new releases (\RGNew{} against \RMLNew).

\begin{table}[t]
\centering
\caption{The comparison by how the movie entered the prompt.}\label{tab:source}
\footnotesize
\setlength{\tabcolsep}{3pt}
\begin{tabular}{lrrrrrrrrrrr}
\toprule
 & & \multicolumn{2}{c}{SD} & \multicolumn{4}{c}{Correlation} & & \multicolumn{3}{c}{Ordering accuracy (\%)} \\
\cmidrule(lr){3-4}\cmidrule(lr){5-8}\cmidrule(lr){10-12}
Source & $N$ & MovieLens & Person & Person & MF & Neighb. & MovieLens & Members & Person & MF & MovieLens \\
\midrule
Broad sample & 60 & 0.61 & 1.48 & 0.84 & 0.58 & 0.62 & 0.54 & 8 & -- & -- & -- \\
Possible recommendation & 445 & 0.44 & 0.99 & 0.64 & 0.52 & 0.52 & 0.21 & 67 & 70 & 65 & 56 \\
New release & 46 & 0.55 & 0.94 & 0.63 & 0.46 & 0.51 & 0.47 & 6 & -- & -- & -- \\
\bottomrule
\end{tabular}
\par\vspace{2pt}\parbox{\linewidth}{\footnotesize\emph{Notes:} SD: standard deviation of the MovieLens prediction and of the guess among the rated movies from that source. Possible recommendations were drawn from the hundred movies the recommender ranked highest for the member among eligible movies the member had not rated. Ordering accuracy is shown when at least 20 members have a pair of differently rated movies.}
\end{table}

\section{How Robust Is the Comparison?}
\label{sec:real}

This section tests whether the comparison depends on the timing of the later rating, on the guess shaping the rating, on members copying the recommender's displayed number, on which guesses get rated, or on how the machines were built. Table~\ref{tab:real} collects the evidence.

\begin{table}[t]
\centering
\caption{The comparison across timing, sample definitions, exposure and selection.}\label{tab:real}
\footnotesize
\setlength{\tabcolsep}{3pt}
\begin{tabular}{lrrrrrrrrr}
\toprule
 & & \multicolumn{4}{c}{Correlation} & \multicolumn{4}{c}{Ordering accuracy (\%)} \\
\cmidrule(lr){3-6}\cmidrule(lr){7-10}
Sample & $N$ & Person & MF & Neighb. & MovieLens & Members & Person & MF & MovieLens \\
\midrule
\multicolumn{10}{l}{\emph{A. Minimum gap between guess and rating}} \\
\quad 3 hours & 592 & 0.71 & 0.56 & 0.57 & 0.35 & 80 & 72 & 62 & 56 \\
\quad 1 day (main) & 551 & 0.69 & 0.54 & 0.56 & 0.34 & 77 & 72 & 64 & 56 \\
\quad 7 days & 479 & 0.66 & 0.56 & 0.56 & 0.37 & 69 & 70 & 64 & 54 \\
\quad 30 days & 355 & 0.69 & 0.55 & 0.55 & 0.44 & 50 & 67 & 69 & 56 \\
\quad 90 days & 213 & 0.59 & 0.50 & 0.46 & 0.43 & 27 & 56 & 66 & 52 \\
\quad 180 days & 92 & 0.59 & 0.37 & 0.37 & 0.33 & 12 & -- & -- & -- \\
\addlinespace[3pt]\multicolumn{10}{l}{\emph{B. Fixed follow-up: guesses followable for $H$ days, rated within $H$ days}} \\
\quad $H$ = 90 days & 297 & 0.71 & 0.54 & 0.59 & 0.31 & 35 & 81 & 66 & 55 \\
\quad $H$ = 180 days & 344 & 0.66 & 0.54 & 0.58 & 0.32 & 40 & 73 & 70 & 61 \\
\addlinespace[3pt]\multicolumn{10}{l}{\emph{C. Sample definition, repeated prompts and model timing}} \\
\quad Including known earlier experience & 570 & 0.70 & 0.55 & 0.56 & 0.35 & 79 & 72 & 64 & 56 \\
\quad No second guess before the rating & 453 & 0.70 & 0.59 & 0.60 & 0.40 & 70 & 75 & 67 & 63 \\
\quad Not prompted again before the rating & 249 & 0.75 & 0.63 & 0.64 & 0.38 & 28 & 72 & 68 & 67 \\
\quad Only one recorded guess for this pair & 451 & 0.70 & 0.59 & 0.60 & 0.39 & 70 & 75 & 67 & 63 \\
\quad Guesses from May 2023 on & 437 & 0.68 & 0.54 & 0.55 & 0.37 & 63 & 74 & 65 & 59 \\
\quad Machines without fold-in (monthly snapshot) & 551 & 0.69 & 0.42 & 0.47 & 0.34 & 77 & 72 & 60 & 56 \\
\quad Members first observed during the month & 172 & 0.76 & 0.57 & 0.52 & 0.39 & 19 & -- & -- & -- \\
\quad Members observed before the month & 379 & 0.64 & 0.54 & 0.60 & 0.31 & 61 & 72 & 59 & 54 \\
\addlinespace[3pt]\multicolumn{10}{l}{\emph{D. Other samples}} \\
\quad Not in recorded recommendation lists & 409 & 0.71 & 0.53 & 0.54 & 0.40 & 54 & 74 & 69 & 63 \\
\quad Previously rated, 30+ days before & 238 & 0.67 & 0.70 & 0.73 & 0.69 & 26 & 61 & 66 & 66 \\
\quad Previously rated, 1+ year before & 176 & 0.79 & 0.80 & 0.82 & 0.72 & 20 & 53 & 60 & 64 \\
\addlinespace[3pt]\multicolumn{10}{l}{\emph{E. Reweighted to the self-reported unseen guesses}} \\
\quad Movies with 195+ earlier ratings & 496 & 0.75 & 0.53 & 0.54 & 0.44 & & & & \\
\quad All movies & 551 & 0.74 & 0.50 & 0.50 & 0.37 & & & & \\
\bottomrule
\end{tabular}
\par\vspace{2pt}\parbox{\linewidth}{\footnotesize\emph{Notes:} MF: matrix factorization; Neighb.: item neighborhood. Ordering accuracy is shown when at least 20 members have a pair of differently rated movies. Panel C, first row: the original definition, which keeps movies the member had rated or reported seeing before the guess. For previously rated movies the outcome is the rating entered before the guess; no model built here saw it, while the MovieLens prediction may have. Panel E weights each rated movie by the inverse of its estimated probability of being rated (Appendix~\ref{app:selection}); the first row restricts the target to movies with a realistic chance of being rated, where the weights balance the observed covariates, and the second keeps all guesses, where they do not balance movie popularity.}
\end{table}

\subsection{Timing, follow-up and repeated prompts}

If members simply repeated the number they had given, agreement would be highest for ratings entered soon after the guess and would fade as the guess was forgotten. The correlation advantage does not fade. Requiring at least thirty days between guess and rating leaves \NMonth{} pairs, and the guess correlates \RGMonth{} with the later rating, against \RMFMonth{} for matrix factorization and \RMLMonth{} for MovieLens. At ninety days the figures are \RGQuarter{}, \RMFQuarter{} and \RMLQuarter{}, and at six months \RGHalf{}, \RMFHalf{} and \RMLHalf{}.

Ordering accuracy behaves differently over long gaps. It needs members with at least two differently rated movies in the same window, and their number falls quickly with the gap. With a minimum gap of seven days the guess still leads (\ConcGWeek\% against \ConcMFWeek\% for matrix factorization, \ConcKNWeek\% for the item neighborhood and \ConcMLWeek\% for MovieLens). At thirty days (\NMonthMembers{} members) the guess and matrix factorization order movies about equally well (\ConcGMonth\% and \ConcMFMonth\%), and at ninety days (\NQuarterMembers{} members) the trained recommenders order them better than the guess (\ConcMFQuarter\% for matrix factorization and \ConcKNQuarter\% for the item neighborhood, against \ConcGQuarter\%). In ordering, the person's lead is largest for ratings entered soon after the guess and gone for ratings entered months later, while the advantage in correlation persists. Pairs in the main measure can come from guesses made months apart; restricted to pairs of movies guessed within seven days of each other (\NearPairs{} pairs from \NearMembers{} members), the guess orders \ConcGNear\% correctly, against \ConcMFNear\% for matrix factorization, \ConcKNNear\% for the item neighborhood and \ConcMLNear\% for MovieLens.

Guesses made late in the study could be followed for less time than early ones. Fixing the follow-up window removes that difference. Among guesses that could be followed for at least ninety days, \NHNinety{} were rated within ninety days, and on these the guess correlates \RGHNinety{} with the later rating, against \RMFHNinety{} for matrix factorization, \RKNHNinety{} for the item neighborhood and \RMLHNinety{} for MovieLens; with a 180-day window the figures are \RGHHalf{}, \RMFHHalf{}, \RKNHHalf{} and \RMLHHalf{} (\NHHalf{} pairs).

Prompts could also have reminded members of their guess. For \NReprompt{} of the \NLater{} pairs, the movie appeared in a prompt again before the rating, and for \NReguess{} of them the member gave a second guess. Dropping the pairs with a second guess, the guess correlates \RGNoReguess{} with the later rating, against \RMFNoReguess{} and \RMLNoReguess{}; dropping every pair prompted again, \RGNoReprompt{} against \RMFNoReprompt{} and \RMLNoReprompt{}. None of these checks shows the fading that an echo of a recently stated or repeated number would produce.

\subsection{Movies members had rated before}

A sharper test reverses the order of events. For \NForgot{} guesses, the member had rated the movie at least thirty days \emph{before} answering ``not seen'' (a median of \ForgotYears{} years before). Their guess cannot have influenced a rating that already existed. It correlates \RGForgot{} with that rating, and \RGForgotYear{} when the rating is a year old or more. The guess therefore carries information about the member's taste on its own, without any help from the guess shaping the outcome.

On these movies the machines do as well as the person or better (\RMFForgot{} for matrix factorization, \RKNForgot{} for the item neighborhood, \RMLForgot{} for MovieLens). Previously rated movies differ from the main sample in several ways at once. They are older (a median of \AgeForgot{} years against \AgeLater{}), more widely rated (a median of \NMovieForgot{} prior ratings against \NMovieLater{}), and rated by the platform's heaviest users (a median of \NUserForgot{} past ratings against \NUserLater{}); \BroadForgot\% come from the broad sample of popular and highly rated titles, against \BroadLater\% of the main sample. These are the conditions under which a recommender knows most about both the movie and the member, which is where the account in Section~\ref{sec:theory} expects the person's edge to end.

\subsection{Did members copy the recommender?}

MovieLens shows its predicted ratings throughout the site, so a member might have seen the number before answering. Two checks show that copying does not drive the result. First, restricted to the \NNoExp{} pairs whose movie had not appeared in the member's recorded recommendation lists before the guess, the guess correlates \RGNoExp{} with the later rating, against \RMFNoExp{} and \RMLNoExp{}. Second, wherever a member might have seen the number, guesses rarely copy it. Across all guesses, the guess equals the MovieLens prediction rounded to the half star \CopyEqual\% of the time, against \CopyShuffled\% when each member's guesses are matched at random to the predictions for their other prompts; the average distance between the two is \GapML{} stars, and \GapOneStar\% of guesses differ from the prediction by a star or more.

\subsection{Which guesses get rated}

Members rate the movies they choose to watch, and they choose what they expect to like. Of the self-reported unseen movies guessed at five stars, \WatchHigh\% were later rated, against \WatchLow\% of those guessed at one star; and the surer a member was, the more likely a rating was to follow (\WatchCertLow\% at certainty one, \WatchCertHigh\% at five). Scoring every forecast on the same pairs keeps the selected set the same for all of them, but the set could still favor one forecaster.

We therefore reweight the rated pairs to resemble the guesses that were not rated. A logistic model predicts whether a guess is followed by a rating from the guess, certainty, the machine forecasts, how many ratings the member and the movie have, the prompt source and how long the guess could be followed, and each rated pair is weighted by the inverse of that probability (Appendix~\ref{app:selection}). The weights recover accuracy on the target guesses when, given these variables, whether a rating is recorded does not depend on the rating itself. Guesses about obscure movies are almost never followed by a rating, so we target the \OvN{} guesses about movies with at least \OvMin{} earlier ratings, the tenth percentile among rated movies, a cut that uses no outcomes. On this target the weights balance every observed covariate to within \OvBalMax{} standard deviations. The guess correlates \OvG{} with the later rating, against \OvMF{} for matrix factorization, \OvKN{} for the item neighborhood and \OvML{} for MovieLens. The 95\% intervals for the differences from matrix factorization and MovieLens are \OvDiffCIMF{} and \OvDiffCIML{}, refitting the selection model in each draw (effective sample size \OvEss{} of \OvRated). Weighting toward all \NIPWPop{} self-reported unseen guesses gives similar correlations (\IPWG{} against \IPWMF, \IPWKN{} and \IPWML; Appendix~\ref{app:selection}).

\subsection{How the machines were built}

The trained machines combine a movie side, fitted at the start of each month, with a member side that folds in every rating the member entered before the guess. Folding in matters: of the \NLater{} pairs, \NNewMem{} come from members whose first recorded rating came after the month's cutoff and who had entered a median of \NewMemHistMedian{} ratings by the time of the guess. A monthly snapshot knows nothing about these members; without folding in, matrix factorization correlates \RMFSnap{} with the later rating and the item neighborhood \RKNSnap{}, against \RMF{} and \RKN{} with it. The guess leads among these members (\RGNewMem{} against \RMFNewMem{} for matrix factorization, \RKNNewMem{} for the item neighborhood and \RMLNewMem{} for MovieLens) and among members observed before the month (\RGEstab{} against \RMFEstab, \RKNEstab{} and \RMLEstab), where the item neighborhood comes closest. In squared error, the estimated advantage over matrix factorization is \NewAdvMF{} among the newly observed members and \KnownAdvMF{} among the others (95\% interval for the difference \NewDiffCIMF{}; against the item neighborhood, \NewDiffCIKN): the advantage is larger where the machines know the member least, as the account in Section~\ref{sec:theory} expects.

Three further choices could handicap the machines. First, the hyperparameters were tuned on ratings from March and April 2023, which overlap the first guesses; restricted to guesses made from May 2023 on (\NAfterTune{} pairs), the guess correlates \RGAfterTune{} with the later rating, against \RMFAfterTune{} and \RMLAfterTune{}. Second, the movie side ages over the month. The person's advantage in squared error does not grow detectably with the days since the cutoff (slope \StaleSlope, standard error \StaleSE); in the first half of the month the guess correlates \RGEarly{} with the later rating and matrix factorization \RMFEarly, and in the second half \RGLate{} and \RMFLate. Third, stochastic training could have been unlucky; retrained with two other seeds, matrix factorization correlates \MFSeedRange{} with the later rating and orders \MFSeedConcRange\% of pairs correctly. The results also do not hinge on single members or single movies (Appendix~\ref{app:robust}).

\subsection{Repeat ratings as a reference}

Ratings are noisy. Across the \NRetest{} pairs in which a member rated the same movie twice, a median of \CeilGapYears{} years apart, the two ratings correlate \CeilR{}, agree exactly \CeilExact\% of the time, and order a member's movies the same way \CeilConc\% of the time. This agreement is a reference, not a ceiling: a forecast of a member's stable taste can agree with a later rating more closely than one noisy rating entered years earlier. The guess (\ConcG\%) and the trained recommenders (\ConcMF\% to \ConcMachMax\%) all order movies well above a coin flip and below the agreement between a member's own ratings of the same movie.

\section{Where Do People Know More?}
\label{sec:where}

\subsection{The edge across movies and members}

\begin{figure}[t]
  \centering
  \includegraphics[width=0.9\linewidth]{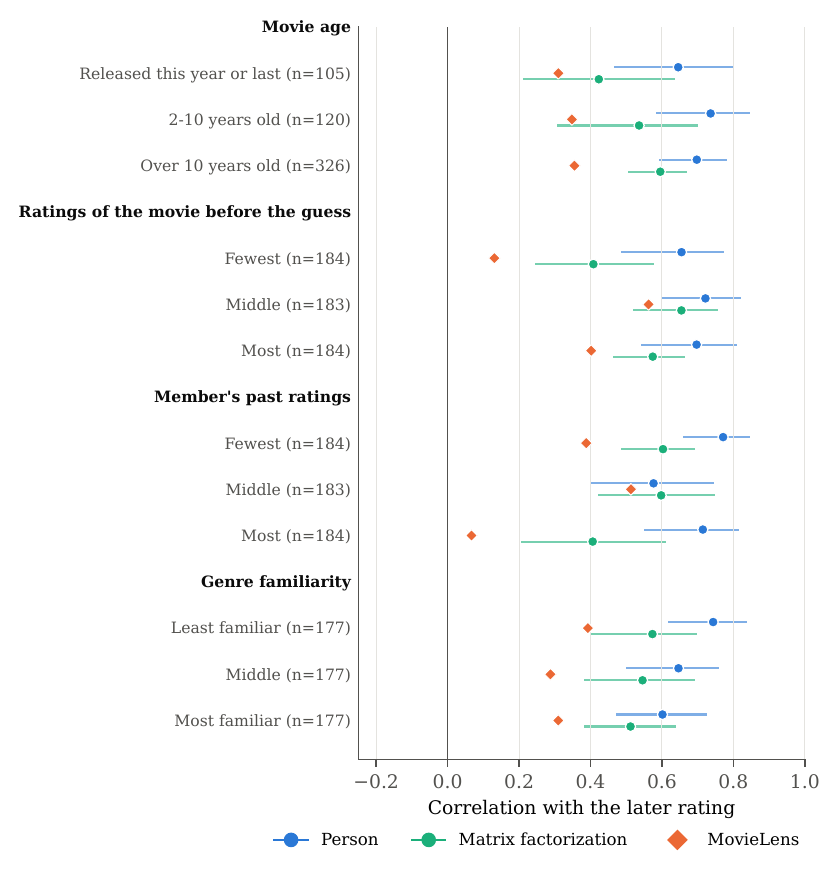}
  \caption{Correlation with the later rating, by group, with 95\% intervals from resampling members. Movie age is measured from the release year to the year of the guess; the other splits are thirds of the rated movies.}
  \label{fig:where}
\end{figure}

Figure~\ref{fig:where} splits the main sample four ways: by the movie's age, by how many ratings the movie had before the guess, by how many ratings the member had entered, and by how familiar the movie's genres were to the member (the share of the member's earlier ratings that share a genre with it). The guess correlates more strongly with the later rating than matrix factorization does in \HetLead{} of the \HetCells{} groups, and in \HetSig{} of them the 95\% interval for the difference excludes zero. The guess's correlation ranges from \HetGMin{} to \HetGMax{} across groups, and matrix factorization's from \HetMMin{} to \HetMMax{}.

The theory in Section~\ref{sec:theory} predicts that the person's edge should be largest where the machine's evidence about the match is thin. Because the four features are correlated, we weigh them together: we regress the difference in squared error between matrix factorization and the guess (positive when the guess is closer) on the four features at once, standardized, with errors clustered by member. At the average of the features the guess's squared error is \WRegConst{} smaller. Table~\ref{tab:hetreg} in the appendix repeats the regression against each of the other machines.

\begin{itemize}
  \item \emph{Thin data about the movie.} The edge shrinks as the movie has more prior ratings (coefficient \WRegM{} per standard deviation, standard error \WRegMSE, $p = \WRegMP$). Among the third of movies with the fewest prior ratings, the guess correlates \RGFewest{} with the later rating, matrix factorization \RMFewest{} and the MovieLens prediction \RMLFewest.
  \item \emph{Unfamiliar territory.} The edge also shrinks as the movie's genres become more familiar to the member (coefficient $-$\WRegFamAbs, standard error \WRegFamSE, $p = \WRegFamP$). In the least familiar third the guess correlates \RGUnfam{} with the later rating and matrix factorization \RMUnfam; in the most familiar third, \RGFam{} and \RMFam.
  \item \emph{Thin data about the member.} The edge leans toward members with fewer past ratings (coefficient \WRegU, standard error \WRegUSE), in line with the larger estimated advantage among members first observed during the month (Section~\ref{sec:real}).
  \item \emph{New releases.} The edge is not detectably larger for movies released in the year of the guess or the year before (coefficient \WRegNew, standard error \WRegNewSE). For a new release the person and the machine are both short of information, the person about the experience itself and the machine about the movie.
\end{itemize}

The coefficients on the movie's ratings, the member's ratings and genre familiarity are negative against every machine (Table~\ref{tab:hetreg}). The pattern matches the account: relative to the machine, the person knows most about the movies and the parts of their taste that the rating data show least.

\subsection{Do people know when they know?}

Members also said how sure they were of each guess. \citet[Appendix~A]{aridor2022informational} show that the gap between a member's expected and later rating grows with the member's stated uncertainty, and that members are surer about extreme expectations; both patterns hold here (certainty correlates \CertExtreme{} with the distance of the guess from three stars). Our question is how certainty bears on the comparison with the machine.

For guesses at certainty four or five (\NSure{} pairs), the guess correlates \RGSure{} with the later rating and matrix factorization \RMSure{} (95\% interval for the difference \DiffSureCI). For guesses at certainty one or two (\NUnsure{} pairs), the figures are \RGUnsure{} and \RMUnsure{} (\DiffUnsureCI). At certainty five (\NFive{} pairs), the guess misses the later rating by \MAEGFive{} stars on average and matrix factorization by \MAEMFive{} stars. That large gap reflects extremity more than certainty. Members are surest about the movies they expect to love or dislike, and the machines rarely forecast extremes. Regressing the person's advantage in squared error on certainty and on the guess's distance from three stars, the advantage grows with that distance (coefficient \LeadExtreme, standard error \LeadExtremeSE) but not with certainty (certainty five against one: \LeadFive, standard error \LeadFiveSE). Certainty tells a platform which guesses are extreme more than which guesses to trust over the machine; Section~\ref{sec:value} measures what it adds to a forecast.

Certainty is not free of bias. Holding the guess fixed, the later rating falls \CertBiasFiveAbs{} stars further below it at certainty five than at certainty one (standard error \CertBiasFiveSE): the surest members are somewhat overoptimistic.

\section{What Does the Guess Add to a Recommender?}
\label{sec:value}

\subsection{What the guess carries beyond the machine information}

If the guess merely restated what a recommender already knows, asking would be pointless. We predict each of the \NGuesses{} guesses from the machine information: every machine forecast, the movie's and the member's rating counts, the movie's age and genres, the member's familiarity with those genres, the prompt source and the member's recommendation lists before the guess. A gradient-boosted model fitted with members held out explains \GuessMachRsqPct\% of the variance in guesses. Among the rated movies, the part of the guess this model predicts correlates \RGuessHat{} with the later rating, and the part it does not predict still correlates \RUnexpl{}. Residualizing the later rating on the same machine information as well, out of sample, the two residuals correlate \RPartial{}. The machine information leaves most of the guess unexplained, and the unexplained part carries signal about the later rating.

\subsection{The held-out gain from asking}

Table~\ref{tab:combine} fits forecasts of the later rating from the machine information, then adds the guess, certainty and their interaction, using ten-fold cross-validation that holds each member's movies out together and the same folds for every model. Machine information alone is the stacked machine of Table~\ref{tab:main}: a correlation of \AblMachR{} with the later rating ($R^2 = \AblMachRsq$). Adding the guess raises it to \AblGuessR{} ($R^2 = \AblGuessRsq$), cutting the mean squared error by \PairGuessMSE{} (95\% interval \PairGuessMSECI), \PairGuessMSEShare\% of its machine-only level. Ordering accuracy moves from \AblMachConc\% to \AblGuessConc\% (95\% interval for the change \PairGuessConcCI{} points), consistent with Section~\ref{sec:main}. Gradient boosting gives the same picture (correlation \AblGbmMachR{} without the guess, \AblGbmAllR{} with it). The gain does not depend on training with outcomes from later in the study: fitting both models on the \FwdTrain{} rated guesses made before \FwdSplit{} and evaluating them on the \FwdTest{} later ones, adding the guess cuts the mean squared error from \FwdMSEMach{} to \FwdMSEGuess{} (reduction \FwdGain, interval \FwdGainCI).

The expected rating carries the gain. Certainty adds \PairCertMSE{} (95\% interval \PairCertMSECI), and letting certainty change the weight on the guess adds nothing detectable (\PairIntMSE, interval \PairIntMSECI).

The reverse comparison is asymmetric. The guess and certainty alone reach $R^2 = \AblPersRsq$, more than all the machine information together. Adding all the machine information to them reduces the error by \PairPersMSE{} (\PairPersMSECI), a gain indistinguishable from zero out of sample, although matrix factorization keeps a substantial weight in the regression of Section~\ref{sec:main}. The guess adds far more to the machines than the machines add to the guess.

\begin{table}[t]
\centering
\caption{What the guess adds to the machine information, out of sample.}\label{tab:combine}
\footnotesize
\setlength{\tabcolsep}{3.4pt}
\begin{tabular}{lrrrrrc}
\toprule
 & & & Ordering & & \multicolumn{2}{c}{Reduction in MSE} \\
\cmidrule(lr){6-7}
Inputs & Correlation & $R^2$ & accuracy (\%) & MSE & Estimate & 95\% CI \\
\midrule
\multicolumn{7}{l}{\emph{A. Linear models; each row adds to the one above}} \\
\quad Machine information & 0.59 & 0.34 & 66.2 & 0.921 & &  \\
\quad + guess & 0.72 & 0.51 & 70.7 & 0.685 & 0.236 & [0.122, 0.363] \\
\quad + certainty & 0.73 & 0.53 & 68.2 & 0.664 & 0.022 & [$-$0.001, 0.046] \\
\quad + guess $\times$ certainty & 0.72 & 0.52 & 71.1 & 0.672 & $-$0.008 & [$-$0.024, 0.007] \\
\addlinespace[3pt]\multicolumn{7}{l}{\emph{B. The member's earlier guesses about other movies (linear)}} \\
\quad Machine + coarse summaries of earlier guesses & 0.60 & 0.35 & 67.0 & 0.907 & 0.014 & [$-$0.027, 0.057] \\
\quad \quad + current guess & 0.71 & 0.51 & 69.1 & 0.694 & 0.213 & [0.106, 0.339] \\
\quad Machine + coarse and similarity-weighted summaries & 0.61 & 0.37 & 67.0 & 0.885 & 0.036 & [$-$0.019, 0.092] \\
\quad \quad + current guess & 0.71 & 0.50 & 70.0 & 0.702 & 0.183 & [0.085, 0.295] \\
\addlinespace[3pt]\multicolumn{7}{l}{\emph{C. Without the machine}} \\
\quad Guess, certainty and their interaction & 0.70 & 0.49 & 72.4 & 0.721 & &  \\
\quad Guess alone, rescaled & 0.68 & 0.47 & 71.6 & 0.748 & &  \\
\addlinespace[3pt]\multicolumn{7}{l}{\emph{D. Gradient boosting}} \\
\quad Machine information & 0.56 & 0.30 & 63.6 & 0.978 & &  \\
\quad Machine + guess + certainty & 0.70 & 0.49 & 64.3 & 0.719 & 0.259 & [0.144, 0.401] \\
\bottomrule
\end{tabular}
\par\vspace{2pt}\parbox{\linewidth}{\footnotesize\emph{Notes:} Ten-fold cross-validation with each member's movies held out together; every model uses the same folds, and missing feature values are filled from the training folds. Machine information alone is the stacked machine of Table~\ref{tab:main}. MSE: mean squared error of the out-of-fold forecast. The reduction in MSE compares each row with the row above it (Panel A), with machine information alone (Panel B, rows 1 and 3), with the row above it (Panel B, rows 2 and 4) and with machine information alone (Panel D); intervals resample members, conditional on the out-of-fold forecasts. Machine information: the MovieLens prediction, matrix factorization, the item neighborhood, the user + movie baseline, movie and user means, rating counts, the movie's age and genres, the member's genre familiarity, the prompt source and the member's recommendation lists before the guess. Coarse summaries of earlier guesses: over the member's guesses about other movies made before this one, the average guess and the average gaps between guess and the MovieLens and matrix-factorization forecasts. Similarity-weighted summaries: the same gaps (and gaps from the item neighborhood and the baseline) weighted by each earlier movie's collaborative-filtering similarity to the target at the monthly cutoff, the gap for the five most similar earlier movies, and gaps weighted by genre overlap.}
\end{table}

\subsection{Can earlier guesses stand in for the current one?}

A platform could hope to learn what a member knows once and reuse it, for instance by noting that a member's expectations tend to run above or below the machine's, overall or for certain kinds of movies. If so, the member's earlier guesses about other movies would substitute for the guess about the movie at hand. We summarize the member's guesses about other movies made before the current one in two ways. Coarse summaries are the average guess and its average gaps from the MovieLens and matrix-factorization forecasts. Similarity-weighted summaries weight the same gaps by each earlier movie's collaborative-filtering similarity to the target at the monthly cutoff or by its genre overlap with the target, so that a member who expects more than the machine from action movies and less from dramas is represented as such. Of the rated movies, \OthShare\% have at least one earlier guess, with a median of \OthMedian, and \TransferShareSim\% have an earlier guess about a similar movie.

Panel B of Table~\ref{tab:combine} reports the results. Added to the machine information, the coarse summaries reduce the mean squared error by \PairOthMSE{} (95\% interval \PairOthMSECI), and the coarse and similarity-weighted summaries together by \PairTransferMSE{} (\PairTransferMSECI). Neither gain is distinguishable from zero. The current guess, added on top of them, cuts the error by \PairGuessTransferMSE{} (\PairGuessTransferMSECI). The summaries add no detectable value whether the member had made few earlier guesses or many (Appendix~\ref{app:robust}). The value lies in asking about the movie at hand.

\section{Discussion}
\label{sec:discussion}

Asked how much they would like a movie they had not seen, MovieLens members gave forecasts of their later rating that beat the platform's recommender, two recommenders trained on tens of millions of ratings and updated with the member's latest ones, and a stack trained on other members' outcomes. The guess predicts the part of the rating that the member's and the movie's averages leave unexplained, it adds to all the machine information, and summaries of the member's earlier guesses do not replace it. In ordering a member's own movies, the trained recommenders are statistically indistinguishable from the person.

\subsection{Theoretical implications}

\paragraph{Reconciling the crowd view and the view of dispersed knowledge.} The familiar ranking of human and machine prediction puts the machine first, from clinical judgment \citep{grove2000} to bail \citep{kleinberg2018} to personality \citep{youyou2015} and other people's tastes \citep{krishnan2008,yeomans2019}. In these comparisons the human forecasts someone else. Here, where the target is one's own rating of a specific movie, the ranking reverses for the rating itself. Neither view of Section~\ref{sec:lit} describes the whole pattern; each holds in a part of it (Table~\ref{tab:predictions}). The crowd view holds where the platform's data are rich. On old, heavily rated movies that members had rated before, the machines match or exceed the person, and in ordering a member's own movies the recommenders that use the member's latest ratings are statistically indistinguishable from the person. This squares with the power of surrogation \citep{gilbert2009}: other people's reactions are an excellent guide to the common part of an experience and to the parts of a person's taste that their record already shows. The view of dispersed knowledge holds for how much a member will like the specific movie asked about. Collaborative filtering estimates the personal match between member and movie, but only from what the member's ratings have revealed. The guess predicts the rating beyond the member's and the movie's averages, it adds to every machine forecast, and simple and similarity-weighted summaries of the member's earlier guesses do not substitute for it. The advantage is larger where the rating data are thinnest, for rarely rated movies and for genres a member seldom rates.

\paragraph{Reversing the information asymmetry.} Platforms are usually the better-informed party: they aggregate the behavior of millions of people, which no single user can see, and recommender design follows from that premise by learning from behavior. For the rating of a specific item, the asymmetry runs the other way. The user knows something the platform does not, and the platform can learn it only by asking. This is the dispersed knowledge of \citet{hayek1945} inside a recommender system. The same asymmetry appears when a crowd forecasts a person's own goals: on an online prediction market, prices overstated how often people achieved the goals they set themselves, and the correction came late and largely from the person \citep{chen2026crowds}. The reversed asymmetry also extends the information systems view of recommendation agents as decision aids \citep{xiao2007}. In that view the consumer receives the agent's output and is influenced by it \citep{adomavicius2013}; our results show the consumer as a source of information the agent lacks.

\paragraph{Unique human knowledge.} \citet{fugener2021} argue that the value of people working alongside AI lies in ``unique human knowledge'', and warn that AI advice can erode it. A member's forecast of their own enjoyment is a clear instance of information that the machines miss. The machine information explains only \GuessMachRsqPct\% of the held-out variation in guesses, and the rest still predicts later ratings; holding all of those features fixed, the guess still correlates \RPartial{} with the later rating. Nor do members copy the platform's number: their guesses match it to the half star about as often as chance would (Section~\ref{sec:real}).

\paragraph{Behavior, experience and expectation.} Recommender systems learn from what people do and from the ratings they give to what they have experienced. A forecast about a specific item not yet experienced is a third kind of data, and it carries information that the other two do not, as \citet{ekstrand2016} anticipated. Because it concerns a concrete item, it can be checked against a later rating, and it differs from general statements of taste.

\subsection{Practical implications: design principles}

Table~\ref{tab:principles} turns the findings into four design principles for preference elicitation and recommender evaluation, each with the evidence behind it and the action it implies for a platform.

\begin{table}[t]
\centering
\caption{Design principles, their evidence and the actions they imply.}\label{tab:principles}
\footnotesize
\begin{tabularx}{\linewidth}{>{\raggedright\arraybackslash}p{3.3cm}>{\raggedright\arraybackslash}X>{\raggedright\arraybackslash}X}
\toprule
Principle & Evidence & Action \\
\midrule
1. Ask for the expected rating of a specific item & Added to all machine information, the guess cuts held-out mean squared error by \PairGuessMSEShare\%, also forward in time; certainty adds \PairCertMSE{} (interval \PairCertMSECI) & Ask how the user expects to rate a named item, on the platform's own rating scale; make a certainty question optional \\
2. Ask about the items being decided on & Coarse and similarity-weighted summaries of earlier guesses add no detectable value; the current guess adds \PairGuessTransferMSE{} beyond them & Ask about the candidates being ranked, when they are being ranked, rather than reusing earlier answers about other items \\
3. Evaluate recommenders on the candidates they re-rank & Among the recommender's top candidates, MovieLens orders \ConcMLRec\% of pairs correctly, the stacked machine \ConcSTRec\% and the guess \ConcGRec\% & Report ordering accuracy within each user's top candidates alongside error across all items \\
4. Give the benchmark what the platform knows & Without folding in members' latest ratings, matrix factorization correlates \RMFSnap{} with the later rating and orders \ConcMFSnap\% of pairs, against \RMF{} and \ConcMF\% with it & Fold each user's latest ratings into the recommender before comparing it with people \\
\bottomrule
\end{tabularx}
\end{table}

\paragraph{Ask for the expected rating, about the items being decided on.} One guess about a specific movie predicts the later rating better than all the machine signals combined; added to those signals, it cuts the error of a held-out forecast by a quarter, while the signals add little to the guess. The member's certainty adds little, and the surest members are somewhat overoptimistic, so if a platform can ask one question, the expected rating is the one to ask. Summaries of a member's earlier guesses did not replace the current one, so a platform should ask about the items it is deciding on. Asking costs attention, and most MovieLens prompts went unanswered. The design question is therefore how to ask in a way people want to answer, at the moment a decision is being made.

\paragraph{Evaluate recommenders where they operate, with what the platform knows.} Most rated movies came from the hundred the recommender ranked highest for the member, and among these the production system's predictions hardly separated the movies the member would like from those they would not. Offline accuracy measured across all items can hide this, because it rewards separating good movies from bad ones, which the crowd does well. Comparisons of people with recommenders also depend on what the recommender is allowed to know. Monthly snapshots that missed members' latest ratings understated the trained recommenders, most for members new to the platform, and made the person's lead in ordering look larger than it is. A fair benchmark folds in everything the platform has recorded up to the moment of the forecast.

\subsection{Limitations and future work}

Our evidence comes from the guesses that were followed by a recorded rating: \NLater{} pairs from \NLaterUsers{} members, \LabelShare\% of all guesses, on one platform whose members are enthusiasts who rate many movies and who chose to answer the prompt. Watching and rating were the members' choices, and our reweighting assumes that, given the observed variables, whether a rating is recorded does not depend on the rating itself; it also covers only movies with a realistic chance of being rated. A design in which people forecast and then experience assigned items would remove that assumption and extend the comparison to other platforms, items and users.

The predictive signal in the guess may combine private self-knowledge with other sources: information about the movie that our benchmarks omit, a member's use of the rating scale, an earlier viewing that the data do not record (the release records when a rating was entered, not when the movie was watched), a predicted rating the member saw elsewhere on the site, or asking itself, by drawing attention to a movie \citep{aridor2022informational} or by pulling the rating toward the guess. The persistence of the correlation over months, the checks on repeated prompts and the accuracy of guesses about movies rated long before argue against the last of these. A field experiment that randomizes whether a platform asks and whether it uses the answer would separate these sources and would measure what our offline comparison does not: whether asking improves the recommendations people act on.

Two further directions follow. First, we tested coarse and similarity-weighted summaries of a member's earlier guesses; richer representations could show whether some of what a member knows can be reused, which would reduce how often a platform needs to ask. Second, our recommenders are standard, well-tuned members of the two main families of collaborative filtering, with a movie side refitted monthly, and the production system is a black box. Sequential, deep and language-model recommenders are the natural next benchmark.

\section{Conclusion}

Platforms must decide whether to learn what people want only from what they have done, or also by asking. Theories of statistical prediction imply that asking is redundant; theories of dispersed knowledge imply that people hold information the data lack. In MovieLens, what people say they expect from a specific movie they report not having seen predicts their later rating better than a production recommender, better than models trained on tens of millions of ratings and updated with the member's latest ones, and better than a stack trained on other members' outcomes, and it adds to the forecasts and features these machines use. The crowd's ratings reveal which movies are good and much about which suit a member. The member knows something more about the movie in front of them, and a recommender that asks can learn it.

\bibliographystyle{plainnat}
\bibliography{references}

\appendix
\section{Data Construction Details}
\label{app:data}

\paragraph{Guesses.} A guess is a prompt answered ``not seen'' with an expected rating between half a star and five stars. The prompt log contains \NNotSeen{} such answers. When a member was asked about the same movie more than once (\NRepeated{} member--movie pairs), we keep the first answer, so every guess precedes any later prompt about the same movie. Members who were asked twice, at least thirty days apart, gave guesses that correlate \RGuessRetest{} with each other (\NGuessRetest{} pairs).

\paragraph{Ratings.} The release contains the complete rating histories of the members who answered at least one prompt, plus the rating histories of other members who were shown prompts. A rating of $-1$ in the histories is a non-rating code and is dropped. A member can rate the same movie more than once. The later rating is the first rating entered after the guess, used only if it came at least one day after the guess; if the first rating came sooner, the guess has no later rating. For previously rated movies the outcome is the most recent rating entered at least thirty days before the guess.

\paragraph{Known earlier experience.} The main sample excludes guesses whose movie the member had rated at any time before the guess (\NExclPriorRating{} of the guesses with a later rating) and guesses whose movie the member later marked as seen in another prompt, reporting a viewing date before the day of the guess (\NExclPriorView{}). Viewing dates are reported only in such ``seen'' answers; for the other later ratings the release records when the rating was entered, not when the movie was watched.

\paragraph{Clock alignment.} The prompt log stores Coordinated Universal Time and the rating histories store US Central time. When a member answers ``seen'' in a prompt and rates the movie there, the same rating is written to the history; before alignment, the two records differ by a median of almost exactly five hours (summer) or six hours (winter). After converting the histories from US Central time to Coordinated Universal Time, the two records of the same event typically fall within a minute of each other. All ordering of guesses and ratings uses the aligned times.

\paragraph{Corrections.} Of the \NQuick{} first ratings entered within one day after a guess, \QuickHalfHour{} came within thirty minutes. A rating so soon after the answer ``not seen'' is far more likely to be a correction (the member realizing they had seen the movie) than a viewing. Table~\ref{tab:real} reports results with a three-hour minimum instead.

\section{The Machines}
\label{app:machines}

\paragraph{Training data.} The training pool combines the rating histories in the release with the public MovieLens 32M ratings (ratings through October 2023), \NTrainPool{} ratings after removing every member--movie pair with a guess. MovieLens 32M uses its own member identifiers. Most members in the release also appear in it, with the same ratings at the same second; we identify \NDupAccounts{} such accounts (at least half of an account's ratings match one member in the release exactly) and drop them, \NDupRatings{} ratings, so that each member's history enters training once and the removal of guessed pairs is complete. For each calendar month from March 2023 to May 2024, each model is trained on the latest rating of each member--movie pair entered before the first day of the month, and it predicts every guess made in that month, folding in the member's ratings entered before each guess. The ratings in force at a cutoff range from \NTrainMin{} (March 2023) to \NTrainMax{} (May 2024).

\paragraph{Matrix factorization.} The model predicts $\hat y_{im} = \mu + b_i + b_m + p_i^\top q_m$ and is fitted by stochastic gradient descent on squared error with $\ell_2$ penalties, a learning rate of 0.007 decaying by 8\% per pass, and a separate small penalty on the biases. A movie absent from the training data receives no bias or factor term. To fold in a member, we hold $\mu$, the movie biases $b_m$ and the movie factors $q_m$ at their values from the monthly fit and re-estimate the member's bias and factors from all of the member's ratings entered before the guess, by ridge regression of $y - \mu - b_m$ on $(1, q_m)$ with penalty 10 on the bias and $n \cdot 0.02 + c$ on each factor, where $n$ is the number of ratings. Hyperparameters, including $c$, were chosen on a forecasting task of the same shape as the main analysis: models trained on ratings entered before 1 March 2023 predicted a random sample of \ValN{} first ratings of new movies that study members entered in March and April 2023, folding in each member's ratings entered before each rating. Table~\ref{tab:tune} reports the grid. The chosen model (\MFk{} factors, $c = \MFFoldC$) reached a root mean squared error of \MFValRMSE{} with folding in and \MFValRMSECut{} without it, against \BiasValRMSE{} for the user + movie baseline. Retrained with two other random seeds for the starting factors and the order of the updates, it correlates \MFSeedRange{} with the later rating in the main sample.

\begin{table}[h]
\centering
\caption{Tuning the machines on a forecasting task of the same shape.}\label{tab:tune}
\small
\begin{tabular}{lrrrrr}
\toprule
 & & & & \multicolumn{2}{c}{RMSE} \\
\cmidrule(lr){5-6}
Model & Factors & Penalty & Passes & Snapshot & Fold-in \\
\midrule
User + movie baseline & -- & -- & 15 & 0.933 & 0.841 \\
Matrix factorization & 32 & 0.02 & 20 & 0.912 & 0.762 \\
Matrix factorization & 64 & 0.02 & 20 & 0.911 & 0.759 \\
Matrix factorization & 64 & 0.05 & 20 & 0.920 & 0.788 \\
\bottomrule
\end{tabular}

\vspace{6pt}
\begin{tabular}{lrrrrrr}
\toprule
Item neighborhood & $\lambda = 0$ & $\lambda = 2$ & $\lambda = 5$ & $\lambda = 10$ & $\lambda = 50$ & $\lambda = 200$ \\
\midrule
$k = 20$ & 0.781 & 0.780 & 0.780 & 0.779 & 0.780 & 0.782 \\
$k = 50$ & 0.781 & 0.780 & 0.780 & 0.780 & 0.780 & 0.781 \\
$k = 100$ & 0.785 & 0.784 & 0.784 & 0.784 & 0.784 & 0.785 \\
$k = 200$ & 0.787 & 0.787 & 0.786 & 0.786 & 0.786 & 0.787 \\
\bottomrule
\end{tabular}
\par\vspace{2pt}\parbox{\linewidth}{\footnotesize\emph{Notes:} Trained on ratings before 1 March 2023 and evaluated on first ratings of new movies that study members entered in March and April 2023: a random sample of 20,000 (2,727 members) for matrix factorization and the baseline, and of 5,000 for the item neighborhood. Snapshot forecasts use only the model at the cutoff; fold-in forecasts also re-estimate the member's effects from the member's ratings entered before each rating, with the factor shrinkage chosen on this task ($c = 1$). Item neighborhood: root mean squared error of the fold-in forecast by number of neighbors $k$ and similarity shrinkage $\lambda$; the baseline alone scores 0.864.}
\end{table}

\paragraph{Item neighborhood.} The baseline is $b_{im} = \mu + b_i + b_m$ with $b_m = \sum (y - \mu)/(25 + n_m)$ over the movie's ratings at the cutoff and $b_i = \sum (y - \mu - b_m)/(10 + n_i)$ over the member's ratings entered before the guess \citep{koren2010}. The similarity of two movies is the cosine between their baseline residuals across members, multiplied by $n/(n + \lambda)$, where $n$ is the number of members who rated both. The forecast is $b_{im}$ plus the similarity-weighted average of the member's residuals on the $k$ most similar movies the member has rated, among those with positive similarity. On a random sample of \KnnValN{} pairs from the same tuning task, the grid $k \in \{20, 50, 100, 200\}$ and $\lambda \in \{0, 2, 5, 10, 50, 200\}$ selected $k = \KnnK$ and $\lambda = \KnnShrink$ (root mean squared error \KnnValRMSE{}, against \KnnBaseRMSE{} for the baseline alone).

\paragraph{Stacked machine.} The features are the MovieLens prediction, matrix factorization, the item neighborhood, the user + movie baseline, the movie mean and the user mean; the logged numbers of earlier ratings of the movie and by the member; the movie's age; the share of the member's earlier ratings that share a genre with the movie; indicators for the prompt source and for each genre; whether the movie had appeared in the member's recommendation lists before the guess; and the logged number of lists the member was shown. All features are measured before the guess. The stacked machine is a linear regression fitted with ten-fold cross-validation that holds each member's movies out together; missing values of the movie's age and the member's genre familiarity are filled with medians from the training folds. It is the designated comparator, and Table~\ref{tab:combine} reuses its out-of-fold forecasts as ``machine information''. A gradient-boosted model with the same features and folds correlates \StackGbmR{} with the later rating, against \StackOlsR{} for the linear stack.

\paragraph{The MovieLens prediction.} The release reports, for each prompt, the recommender's predicted rating for the member and movie. The prompt itself showed the title. Section~\ref{sec:real} tests whether members copied the predicted ratings MovieLens displays elsewhere on the site.

\section{Selection Adjustments}
\label{app:selection}

\paragraph{Model.} We model the probability that a self-reported unseen guess, with no known earlier experience and no correction within a day, is followed by a later rating with a logistic regression on the guess and its square, certainty (as categories), the MovieLens, matrix-factorization and neighborhood predictions, the logged numbers of ratings of the member and of the movie, the prompt source, and the logged number of days the guess could be followed and its square. Each rated pair is weighted by the inverse of its fitted probability, and correlations are computed with these weights. Intervals for the weighted differences refit the selection model in each of 500 member-resampled draws.

\paragraph{Targets.} Weighting toward all \NIPWPop{} such guesses (\NIPWMembers{} members) moves the rated pairs' standardized means toward those of all guesses for most covariates (for the guess from \BalGuessRaw{} to \BalGuessW{}, for certainty from \BalCertRaw{} to \BalCertW{}, for the days of follow-up from \BalFollowRaw{} to \BalFollowW{}; all guesses: 0) but leaves the movie's number of earlier ratings unbalanced (\BalMovieRaw{} before weighting, \BalMovieW{} after): many guesses concern obscure movies that are almost never rated afterwards, and the weights we fit cannot stand in for them. The effective sample size is \IPWess{} of \NLater{}, and the largest single weight carries \IPWwMax\% of the total. We therefore restrict the target to the \OvN{} guesses (\OvMembers{} members, \OvMovies{} movies) about movies with at least \OvMin{} earlier ratings, the tenth percentile of that count among rated pairs; \OvRated{} of them were rated. On this target the weights balance every covariate above to within \OvBalMax{} standard deviations, with an effective sample size of \OvEss. The restriction uses no outcomes. The weighted differences from the item neighborhood are \OvDiffCIKN{} on the restricted target and \IPWDiffCIKN{} on all guesses. Capping weights on all guesses at their 99th percentile gives correlations of \IPWTrimG{} for the guess, \IPWTrimMF{} for matrix factorization and \IPWTrimML{} for MovieLens.

\section{Additional Analyses}
\label{app:robust}

\paragraph{Members in the ordering measure.} On the \SplitN{} pairs of the \SplitMembers{} members who later rated at least two movies differently, the guess correlates \SplitRG{} with the later rating and the stacked machine \SplitRST, and the person's mean squared error is lower by \SplitDiffMseST{} (95\% interval \SplitDiffMseCIST). Splitting each forecaster's error into a member-level part, how far it misses the member's average rating, and a within-member part, the difference is \SplitDiffLevelST{} in the member-level part (\SplitDiffLevelCIST) and \SplitDiffWithinST{} within members (\SplitDiffWithinCIST).

\paragraph{Earlier guesses by belief history.} For pairs with at most \HistCut{} earlier guesses by the member, the coarse and similarity-weighted summaries reduce the mean squared error by \HistFewTransfer{} (95\% interval \HistFewTransferCI) and the current guess by a further \HistFewGuess{} (\HistFewGuessCI). For pairs with more earlier guesses, the figures are \HistMoreTransfer{} (\HistMoreTransferCI) and \HistMoreGuess{} (\HistMoreGuessCI).

\paragraph{Asking about some movies only.} We compute, for each rated movie, how much adding the expected rating reduces the squared error of the machine-only forecast, and fit a model that predicts this gain from what the platform knows before asking, on other members only. Asking about the quarter of movies with the highest predicted gain captures \AskQuarter\% of the total gain (95\% interval \AskQuarterCI\%), and asking about half captures \AskHalf\% (\AskHalfCI\%), against 25\% and 50\% under random asking; predicted and realized gains correlate \AskGainCorr.

\paragraph{Resampling movies.} Resampling movies rather than members, the 95\% interval for the difference in correlation between the guess and matrix factorization is \DiffRMovieCIMF{}, and between the guess and the MovieLens prediction \DiffRMovieCIML{}.

\paragraph{Single members.} Dropping each member in turn, the person's ordering advantage over matrix factorization stays between \LooMinMF{} and \LooMaxMF{} percentage points, and over MovieLens between \LooMinML{} and \LooMaxML{}.

\paragraph{Heterogeneity against other machines.} Table~\ref{tab:hetreg} repeats the regression of Section~\ref{sec:where} with the loss gap measured against each machine in turn.

\begin{table}[h]
\centering
\caption{Where the person's edge is larger: the loss gap against each machine.}\label{tab:hetreg}
\small
\begin{tabular}{lrrrr}
\toprule
 & Matrix f. & Neighb. & MovieLens & Stacked \\
\midrule
Constant & 0.32 & 0.32 & 1.01 & 0.05 \\
 & (0.16) & (0.17) & (0.25) & (0.12) \\
New release & $-$0.33 & $-$0.03 & $-$1.45 & 0.30 \\
 & (0.31) & (0.37) & (0.47) & (0.30) \\
Ratings of the movie (log, std.) & $-$0.29 & $-$0.18 & $-$0.75 & $-$0.01 \\
 & (0.14) & (0.17) & (0.23) & (0.13) \\
Member's past ratings (log, std.) & $-$0.36 & $-$0.55 & $-$0.32 & $-$0.22 \\
 & (0.30) & (0.34) & (0.36) & (0.16) \\
Genre familiarity (std.) & $-$0.30 & $-$0.27 & $-$0.40 & $-$0.14 \\
 & (0.17) & (0.19) & (0.23) & (0.11) \\
\bottomrule
\end{tabular}
\par\vspace{2pt}\parbox{\linewidth}{\footnotesize\emph{Notes:} Dependent variable: the machine's squared error minus the guess's squared error (positive when the guess is closer). 531 rated movies with a genre-familiarity measure; standard errors clustered by member in parentheses.}
\end{table}

\end{document}